\documentclass[preprint]{elsarticle}          % twocolumn
\usepackage{amsmath}
\usepackage{amssymb}
\usepackage{epsfig}
\usepackage{psfrag}
\usepackage{tikz}
\usepackage{array}
\usepackage{graphicx}
\usepackage{multirow}
\usepackage{framed}
\usepackage{float}
\usepackage{caption}
\usepackage{balance}
\usepackage{subfig}
\usepackage{tabularx}
\usepackage{url}
\begin{document}

%\journal{Nonlinear Analysis: Real World Applications}

\begin{frontmatter}



\title{Impact of queue feedback on the stability and dynamics of a Rate Control Protocol (RCP) with two delays}
\author{Abuthahir}
\ead{ee12d207@ee.iitm.ac.in}
\author{Gaurav Raina}
\ead{gaurav@ee.iitm.ac.in}
\address{Department of Electrical Engineering, Indian Institute of Technology Madras, Chennai - 600 036, India}
\cortext[GR1]{Correspondence to: A. Abuthahir, Department of Electrical Engineering, IIT Madras,
Chennai 600036, India}
% \author{}
% 
% \address{}

\begin{abstract}
Rate Control Protocol (RCP) uses feedback from routers to assign flows their fair rate. RCP estimates the fair rate using two forms of feedback: rate mismatch and queue size. An outstanding design question for RCP is whether the queue size feedback is useful or not.
To address this, we analyze stability and the bifurcation properties of RCP in both the cases i.e., with and without queue size feedback. The model considers flows with two different round-trip times, operating over a single bottleneck link. By using an exogenous bifurcation parameter, we show that the system loses stability via a Hopf bifurcation and hence we can expect a limit cycle branching from the fixed point. We highlight that the presence of queue feedback can readily destabilize the system. Using Poincar{\`e} normal forms and the center manifold theorem, we show that the Hopf bifurcation is super-critical in the case of RCP without queue feedback. Whereas, in the presence of queue feedback, we show that the system can undergoes a sub-critical Hopf bifurcation for some parameter values. A sub-critical Hopf bifurcation can result in either large amplitude limit cycles or unstable limit cycles, and hence should be avoided in engineering applications. Thus, the presence of queue feedback would create adverse effects on the stability of the emerging limit cycles. In essence, the analytical results of RCP with two delays favor the design choice that uses feedback based only on rate mismatch. The theoretical analysis is validated with numerical computations and some packet level simulations as well. 
\end{abstract}


\end{frontmatter}

\section{Introduction}
Internet congestion control has been an active area of research for several decades \cite{srikant2004}. The Transmission Control Protocol (TCP) is the widely used transport layer protocol which control network congestion in the Internet. It has been shown that the performance of TCP is poor in the future high bandwidth delay network due to its use of implicit feedback and the standard Additive Increase Multiplicative Decrease (AIMD) control law \cite{katabi2002}. This motivates the development of congestion control protocols that relies on more explicit feedback~ \cite{dukkipatircpac,he2017,liu2012XCPHopf,liu2016xcc,mahdian2016xcc,ren2016xcc,Wydrowskimaxnet,Zhangjetmax}. Rate control protocol (RCP) is an explicit congestion control algorithm that received a lot of attention from the research community \cite{abu2015ccdcrcp,baretto2015rcp,balakrishnan2007stability,dukkipatircpac,krv2009,lakshmikantha2008,lei15,mahdian2016xcc,nizar2011,sharma17,sun2012rcp,voice2009maxminrcp,zhong2017}. 

A key motivation for considering RCP, is that it continues to receive attention not only in the currently used host-centric~(IP-based) networks, but also in the future data-centric networking architectures like Named Data Networking~(NDN) \cite{zhangndn2010}. In NDN, there is no IP address, and all data are named with unique names. Moreover, the data can be fetched from multiple sources via multiple paths which makes the implicit signaling mechanism unreliable in NDN \cite{ren2016xcc}. Therefore, researchers focus on employing rate-based RCP-style algorithms in NDN \cite{lei15,mahdian2016xcc,zhong2017}. For example, in \cite{lei15} and \cite{zhong2017}, they combined RCP and the unique features of NDN to develop rate-based schemes for NDN congestion control. Similarly, RCP has attracted interest in wireless and satellite networks. In \cite{baretto2015rcp}, RCP is extended with the support of an algorithm to accurately evaluate the capacity of wireless links. Simulation results in \cite{sun2012rcp} reveal that RCP outperforms TCP in terms of throughput and queue size in the satellite networks. In \cite{sharma17}, the feasibility of implementing RCP with flexible packet processing architectures has been demonstrated, especially in the data center networks. In this paper, our focus will be restricted to one particular design problem that arises in the study of feedback mechanism used by RCP. To expand further, RCP assigns a single fair rate for all the flows traversing the bottleneck link. These rates are computed using control equations at the routers, which in turn employ two forms of feedback: rate mismatch and the queue size. Simulation studies in \cite{krv2009} show that the queue size feedback in RCP, can cause the queue to be less accurately controlled. This conclusion was based on some initial simulations, and more work is needed before one could conclude that the queue size terms should be dropped from the protocol definition. In this paper, we focus on the proportionally fair variant of RCP which was introduced in~\citep{krv2009}. The model considers single bottleneck network carrying flows with two different round-trip times.

An RCP router utilizes a field in the packet header to convey the fair share rate at which the flows can send data into the network. However, the feedback about the fair rate to end-systems is not instantaneous. Therefore, RCP works like a closed-loop control system with feedback delays. In general, the stability of a closed-loop system is sensitive to feedback delays, which normally necessitates a detailed stability analysis.
% It is also important to make sure that the system converges quickly to a stable equilibrium. Therefore, we explore the impact of queue feedback on the local stability and the convergence properties of RCP. 
Local stability analysis retains only the linear component and ignores all higher order terms of the non-linear system before addressing the issue of stability. So, it looks appealing to have an analytical methodology which may allow us to capture the impact of some non-linear terms while performing a Taylor expansion of the non-linear system about its equilibrium. Local bifurcation theory is one such methodology \cite{hassard1981}.  A comprehensive understanding of local bifurcation phenomena may help yield insights into the role played by different forms of feedback in RCP. Moreover, any congestion control algorithm is not only to ensure local stability of the equilibrium, but also to make sure that any loss of stability, that may happen, results in \emph{stable} limit cycles of small amplitude. We hasten to add that we are not interested in destabilizing our network, but wish to employ the tools offered by local instability analysis to gain some insight into the non-linear properties of both the design options, i.e., with and without queue size feedback. There is considerable interest in analyzing the stability and Hopf bifurcation of the congestion control algorithms \cite{abu2015ccdcrcp,dongnlarwa2013,liu2011nlarwaTCPhopf,pei2018ijbc,voice2009maxminrcp,zhangnlarwa2013}.

We now need to decide which parameter will be used to violate the stability condition and hence act as the bifurcation parameter. We motivate a non-dimensional exogenous parameter to induce instability. This has various advantages. We need not be concerned with the dimensions of the parameter, and as it is common for both the design choices we can compare the results fairly. By analyzing the roots of the transcendental characteristic equation, we first derive necessary and sufficient conditions for local asymptotic stability. This enables us to determine the stability region in the parameter space. It is then shown that, as the bifurcation parameter varies, the system where feedback is based on both rate mismatch and queue size, readily loses local stability through a Hopf bifurcation~\cite{hassard1981}.
% The rate of convergence analysis could guide us in tuning the parameter values for faster convergence. But, it does not provide any insights on which design choice is desirable. This provides motivation for the analysis which goes beyond local stability and convergence to make progress on this question. 
Then, we investigate the impact of queue feedback on the direction and stability of the emerging limit cycles. To that end, we conduct a detailed Hopf bifurcation analysis for both the design choices. The Appendix contains the necessary calculations to determine the type of Hopf bifurcation and the orbital stability of the bifurcating limit cycles, as local instability just sets in. The theoretical frameworks that we employ to analyze the nature of Hopf bifurcation are the Poincar{\`e} normal forms and the center manifold theorem. We establish that the RCP which uses only rate mismatch feedback would give rise to a super-critical Hopf bifurcation which leads to stable limit cycles of small amplitude. Whereas, in the presence of queue feedback, the system can exhibit a sub-critical Hopf bifurcation, for some parameter values. A sub-critical Hopf bifurcation is undesirable for real engineering systems as a small perturbation around the system equilibrium may give rise to either limit cycles with large amplitude, or unstable limit cycles \cite{strogatz2018}. The numerical analysis tool that we use to validate the theoretical insights is DDE-Biftool (a Matlab package for numerical bifurcation and stability analysis of delay differential equations)~\cite{ddebiftool1},~\cite{ddebiftool2}.  

In summary, the analytical insights of our study tend to favor the design choice where the feedback is based only on rate mismatch. Numerical computations and packet level simulations serve to corroborate the analysis.

The rest of the paper is organized as follows. In Section \ref{sec:rcp2dmodel}, we outline the non-linear fluid model of RCP. We analyze the local asymptotic stability of RCP in Section \ref{sec:rcp2dlsa}. In Section \ref{sec:rcp2dhba}, we conduct a local Hopf bifurcation analysis and highlight the impact of queue feedback on the nature of Hopf bifurcation. In Section \ref{sec:rcp2dconcl}, we conclude with a summary of our contributions, and offer avenues for further research. For ease of exposition, the Hopf bifurcation analysis is contained in an Appendix.

\section{Models}
\label{sec:rcp2dmodel}
The small buffer model of a proportionally fair RCP is governed by the following non-linear delay differential equation~\cite{krv2009}
\begin{equation}
\label{eq:smallbuf_model}
\frac{d}{dt}R_{j}(t)=\dfrac{aR_{j}(t)}{C_{j}\overline{T_{j}}(t)}\Big(C_{j}-y_{j}(t)-b_{j}C_{j}p_{j}\big(y_{j}(t)\big) \Big),
\end{equation}
where
\begin{equation}
y_{j}(t) = \sum\limits_{r:j\in r}x_{r}(t-T_{rj})\nonumber
\end{equation}
is the aggregate load at link $j$ summed over all the routes $r$ passing through link $j$, $R_{j}(t)$ is the rate that RCP maintains for all flows passing through link $j$, $x_{r}(t)$ is the rate on route $r$, $p_{j}(y_{j})$ is the mean queue size at link $j$ when the arriving load is $y_{j}$, $C_{j}$ is the capacity of link $j$, $a$ and $b_{j}$ are non-negative protocol parameters.\\
Here, $T_{j}$ is the average round trip time of packets passing through link $j$ given by
\begin{equation}
\overline{T_{j}} = \dfrac{\sum\limits_{r:j\in r}x_{r}(t)T_{r}}{\sum\limits_{r:j\in r}x_{r}(t)} 
\end{equation}
where,
\begin{equation}
T_{r}=T_{rj} + T_{jr}
\end{equation}
is the sum of the propagation delay from source to link j and the return delay from link $j$ to source.
Here, we assume that the queuing delay is negligible as compared to the propagation delay, which conforms with our assumption of small buffers.
The flow rate $x_{r}(t)$ is given by~\cite{krv2009} 
\begin{equation}
 x_{r}(t) = \left( \sum\limits_{j\in r}R_{j}(t-T_{jr})^{-1}\right)^{-1}
\end{equation}

Now we will model the mean queue size term as follows. Suppose that workload arriving at resource of capacity $C$, over a time period $\tau$ is Gaussian with mean $y_{j}\tau$ and variance $y_{j}\tau\sigma_{j}^2$. The workload present at the queue reflects Brownian motion~\cite{harrison}, with mean under its stationary distribution of 
\begin{equation}
p(y_{j}) = \dfrac{y_{j}\sigma_{j}^2}{2\big(C-y_{j}\big)}.
\end{equation}
The above relation is a simple approximation to mean queue size, where $\sigma_{j}^2$ represents the variability of traffic in the link at a packet level. We assume $\sigma_{j} = 1$, which corresponds to the Poisson arrival of packets of constant size.

For our analysis, we consider the network with single bottleneck link of capacity $C$, carrying two sets of RCP flows with different round trip times; say $\tau_1$ and $\tau_2$. Then the model is given by\\
\begin{equation}
\label{eq:simplified_model}
\frac{d}{dt}R(t)=\dfrac{aR(t)}{C\overline{T}(t)}\Big(C-y(t)-bCp\big(y(t)\big) \Big),
\end{equation}
where
\begin{eqnarray*}
\begin{aligned}
\overline{T} &= (\tau_1+\tau_2)/2\\
y(t) &= R(t-\tau_1)+R(t-\tau_2)
\end{aligned}
\end{eqnarray*}
To model the RCP without queue feedback, the parameter $b$ is set to zero to exclude the queuing term from the RCP model. Then to aim for a particular target link utilisation, say a fraction $\gamma$ of the actual link capacity, $C$ is replaced with  $\gamma C$. Then the model of RCP without queue size feedback is given by
\begin{equation}
\label{eq:simplified_withoutQ_model}
\frac{d}{dt}R(t) = \frac{aR(t)}{\gamma C\overline{T}(t)}\Big(\gamma C-y(t))\Big).
\end{equation}
% To address this question, we derive the conditions for stability and investigate the loss of stability through Hopf bifurcation for both the models.
\newcommand{\myk}{\tilde{a}}
%*************** Local Stability and RoC *************************
\section{Local Stability Analysis}
\label{sec:rcp2dlsa}
In this section, we derive conditions for local stability in terms of model parameters. 
We could induce instability by varying any of the system parameters. However, we prefer not to choose any of the system parameters, but introduce an exogenous non-dimensional parameter $\kappa$ to push the system into the unstable regime.
\subsection{With queue feedback}
Let $R^*$ denote the non-zero equilibrium of equation (\ref{eq:simplified_model}), then
\begin{equation}
\label{eq:butil_equation}
 R^* = \dfrac{C\left(4+b-\sqrt{b^2+8b}\right)}{8}.
\end{equation}
Let $u(t)=R(t)-R^*$ be a perturbation about the equilibrium. Then linearising (\ref{eq:simplified_model}), about the equilibrium, gives
\begin{equation}
\label{eq:linear_equation}
\frac{du(t)}{dt} = -\kappa\myk\big(u(t-\tau_1)+u(t-\tau_2)\big),
\end{equation}
where
\begin{equation}
 \quad \myk = \frac{a}{(\tau_1 + \tau_2)}\left[1+\frac{2R^*}{C}\right].  
\end{equation}
Looking for exponential solutions, the characteristic equation of ~(\ref{eq:linear_equation}) is given by
\begin{equation}
\lambda + \kappa\myk\left(e^{-\lambda\tau_1}+e^{-\lambda\tau_2}\right) = 0,
\label{eq:chareq}
\end{equation}
where $\myk$, $\kappa$, $\tau_1,\tau_2>0$.\\
For the system to be stable, all the roots of the characteristic equation should lie in the left half of the complex plane. 
When the round-trip times are zero, we get $\lambda = -2\kappa\myk < 0$ and hence the system is asymptotically stable. 
However, when $\tau_1, \tau_2 > 0$ the roots may cross the imaginary axis for some values of the system parameters, and hence stability of the system cannot be guaranteed. 
Therefore, the condition for the crossover defines the bounds on the system parameters to maintain stability. To find the critical condition, where this crossover occurs, we substitute $\lambda = \pm i\omega$, $\omega>0$ in (\ref{eq:chareq}). 
Equating the real and imaginary parts, we obtain 
\begin{equation}
\kappa\myk\big(\cos(\omega\tau_1)+\cos(\omega\tau_2)\big) = 0, \label{eq:real0}
\end{equation}
\begin{equation}
\kappa\myk\big(\sin(\omega\tau_1)+\sin(\omega\tau_2)\big) = \omega. \label{eq:img0}
\end{equation}
Solving~(\ref{eq:real0}) and~(\ref{eq:img0}), we get
$$\omega(\tau_1+\tau_2)=(2n+1)\pi,\qquad n=0,1,2,\cdots.$$
We only treat the case $n=0$, which gives $\omega_0=\pi/(\tau_1+\tau_2)$.
We now use the following theorem, stated in \cite{stepan1989}, to get the stability condition.
\newtheorem{theorem}{Theorem}
\begin{theorem}\label{sufficient theorem}
The trivial solution of the scalar delay differential equation
\begin{equation}
\label{eq:stepaneqn1} 
\dot{x}(t) +bx(t-\tau_1)+bx(t-\tau_2)= 0
\end{equation}
is exponentially asymptotically stable if and only if
\begin{equation}
\label{eq:stepaneqn2} 
0 < b < \dfrac{\pi}{2(\tau_1+\tau_2)\cos\left(\dfrac{\pi(\tau_1-\tau_2)}{2(\tau_1+\tau_2)}\right)}.
\end{equation}
\end{theorem}
Comparing (\ref{eq:linear_equation}) with (\ref{eq:stepaneqn1}), the necessary and sufficient condition for the stability of (\ref{eq:linear_equation}) can be written as 
\begin{equation}
\label{eq:stability_relation}
\kappa \myk(\tau_1+\tau_2)\cos\left(\dfrac{\pi(\tau_1-\tau_2)}{2(\tau_1+\tau_2)}\right)<\dfrac{\pi}{2}.
\end{equation}

\begin{figure}[hbtp!]
\centering
\psfrag{x}{\hspace{-16mm}Protocol parameter, $a$}
\psfrag{y}{\hspace{-16mm}Protocol parameter, $b$}
\psfrag{unstble}{\footnotesize{}}
\psfrag{0.79}{\small{$\pi/4$}}
\psfrag{1.57}{\small{$\pi/2$}}
\psfrag{1.18}{\small{$3\pi/8$}}
\psfrag{stbbbbbbbbbbbbbbbb}{\footnotesize{Stable region}}
\includegraphics[trim=0cm 0cm 0cm 2.2cm, clip=true, width=3.25in,height=1.85in]{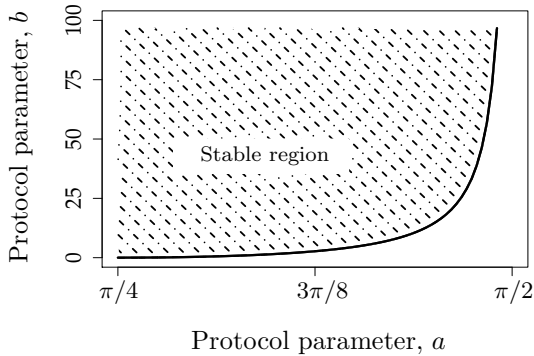}
\vspace{-2mm}
\caption{Stability chart for RCP with queue feedback, highlighting the sufficient condition to ensure local stability for all values of $\tau_1$, $\tau_2 >$ 0.}
%\vspace{-2mm}
\label{fig:ab_stable}
\end{figure}

Substituting the value of $\myk$ in ~(\ref{eq:stability_relation}), we get
\begin{equation}
\label{eq:nsstability_relation}
 \kappa a\left[1+\frac{2R^*}{C}\right]\cos\left(\dfrac{\pi(\tau_1-\tau_2)}{2(\tau_1+\tau_2)}\right)<\pi/2.
\end{equation}
Since $\cos\left(\pi(\tau_1-\tau_2)/2(\tau_1+\tau_2)\right)\in (0,1]$ for all values of $\tau_1$ , $\tau_2 >$ 0, we obtain the sufficient condition as
\begin{equation}
 \kappa a\left[1+\frac{2R^*}{C}\right]<\pi/2.
\end{equation}
For $\kappa=1$, using ~(\ref{eq:butil_equation}), we get 
\begin{equation}
a \left[\dfrac{8+b-\sqrt{b^2+8b}}{4}\right]<\pi/2. 
\end{equation}
See Fig.~\ref{fig:ab_stable} which highlights the sufficient condition to ensure local stability for all values of $\tau_1$, $\tau_2 >$ 0. \\
To show that the system loses local stability through Hopf bifurcation at a critical value of the bifurcation parameter, we need to satisfy the transversality condition of Hopf spectrum which has been outlined in the Appendix.

\subsection{Without queue feedback}
% Now linearizing the equation (\ref{eq:simplified_withoutQ_model}) around its equilibrium, we get
% \begin{equation}
% \label{eq:linear_equation1}
% \frac{du(t)}{dt} = - \kappa\myk\big(u(t-\tau_1)+u(t-\tau_2)\big),
% \end{equation}
% where  $$\quad \myk = \frac{a}{(\tau_1 + \tau_2)} \quad \text{and} \quad R^* = \dfrac{\gamma C}{2}.$$
Proceeding as outlined in the previous sub-section, we get the necessary and sufficient condition for local stability of (\ref{eq:simplified_withoutQ_model}) as
\begin{equation}
\label{eq:nsstability_relation1}
 \kappa a\cos\left(\dfrac{\pi(\tau_1-\tau_2)}{2(\tau_1+\tau_2)}\right)<\pi/2.
\end{equation} 
For $\kappa=1$, we obtain the sufficient condition for stability as
\begin{equation}
  a<\pi/2.
\end{equation}
It is important to highlight that by excluding feedback based on queue size, we can increase the range of the parameter $a$ for which the system is stable.  

We now resort to simulations to study the behaviour of RCP with and without queue feedback. The network being simulated has a single resource or bottleneck link, of capacity $100$ packet per ms, and $100$ sources creating Poisson traffic. Packet level simulations are done using discrete event simulator, which models the behaviour of the RCP network. Equation ~(\ref{eq:butil_equation}) serves as the relation between equilibrium utilisation and the parameter b when the queuing term is present in the RCP definition. for instance, to achieve utilisation of 75\% of link capacity, we need to set $b = 0.166$. To aim for the same utilisation in the case of RCP without queue feedback, we should set b = 0 and $\gamma$ = 0.75. Packet-level traces shown in \ref{fig:Qvsa} illustrate that, in the presence of queue feedback, the system readily loses stability and leads to the emergence of limit cycles. These observations corroborate the results of our stability analysis which establish that the presence of queue size feedback is associated with a smaller choice of the protocol parameter $a$. So as far as the stability is concerned, the observations from simulations favor the design choice of having no queue size feedback in the RCP definition.\\

\newcommand{\hite}{3.16cm}
\begin{figure}[hbtp!]
\psfrag{30.0}{\hspace{2mm}\small{$0$}}
\psfrag{30}{\hspace{2mm}\small{$0$}}
\psfrag{31}{\small{$1000$}}
\psfrag{32}{\small{$2000$}}
\psfrag{32.5}{\small{$2500$}}
\psfrag{35.0}{\small{$5000$}}
\psfrag{0}{\small{$0$}}
\psfrag{250}{\small{$250$}}
\psfrag{500}{\small{$500$}}
\centering
\begin{tabular}{cc} 
	\small{With q(.) feedback} & \small{Without q(.) feedback} \\
	\multicolumn{2}{c}{\small{Queue size (packets)}}\\
\includegraphics[height=\hite, width=0.45\textwidth]{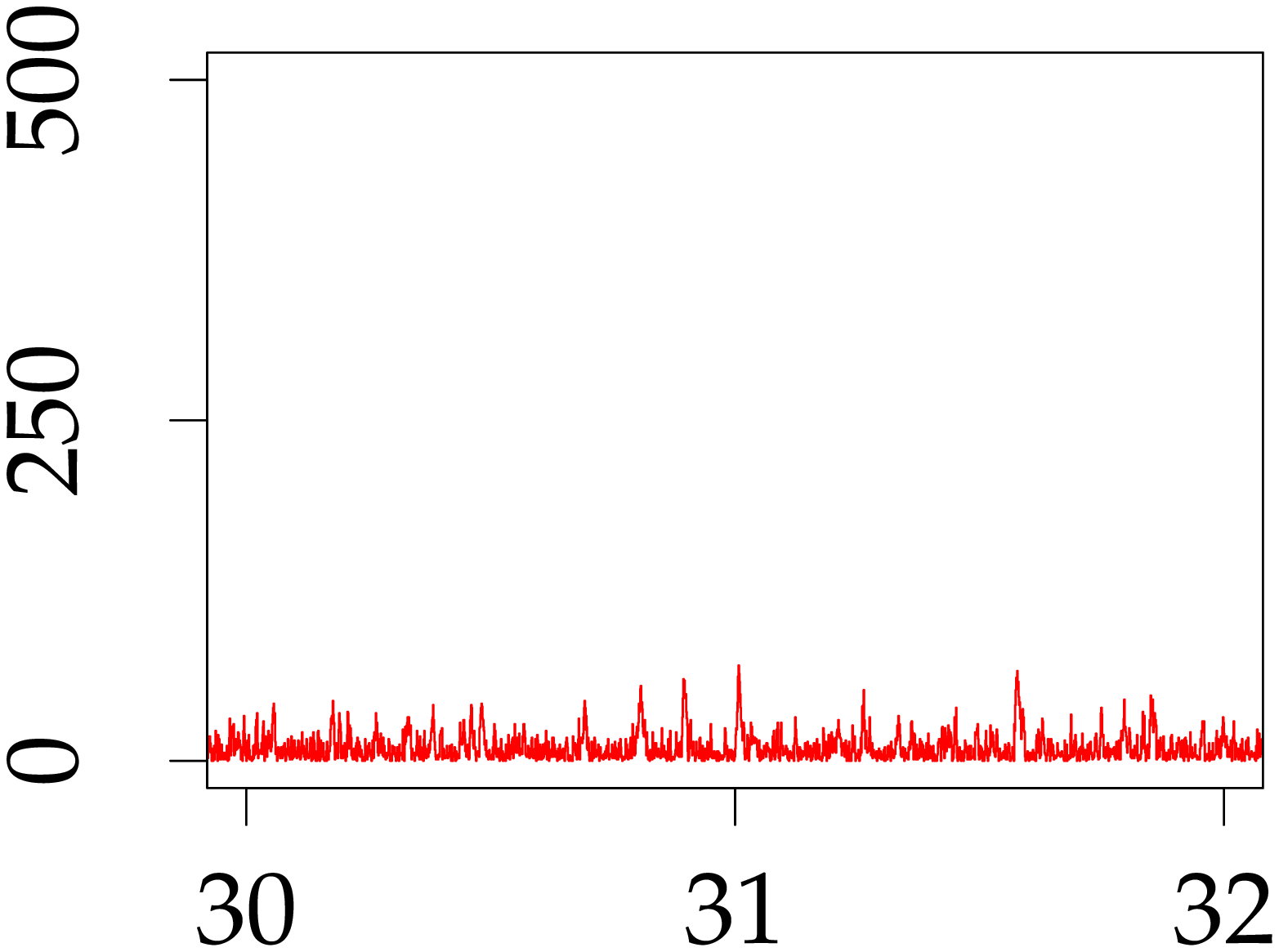} &
\includegraphics[height=\hite, width=0.45\textwidth]{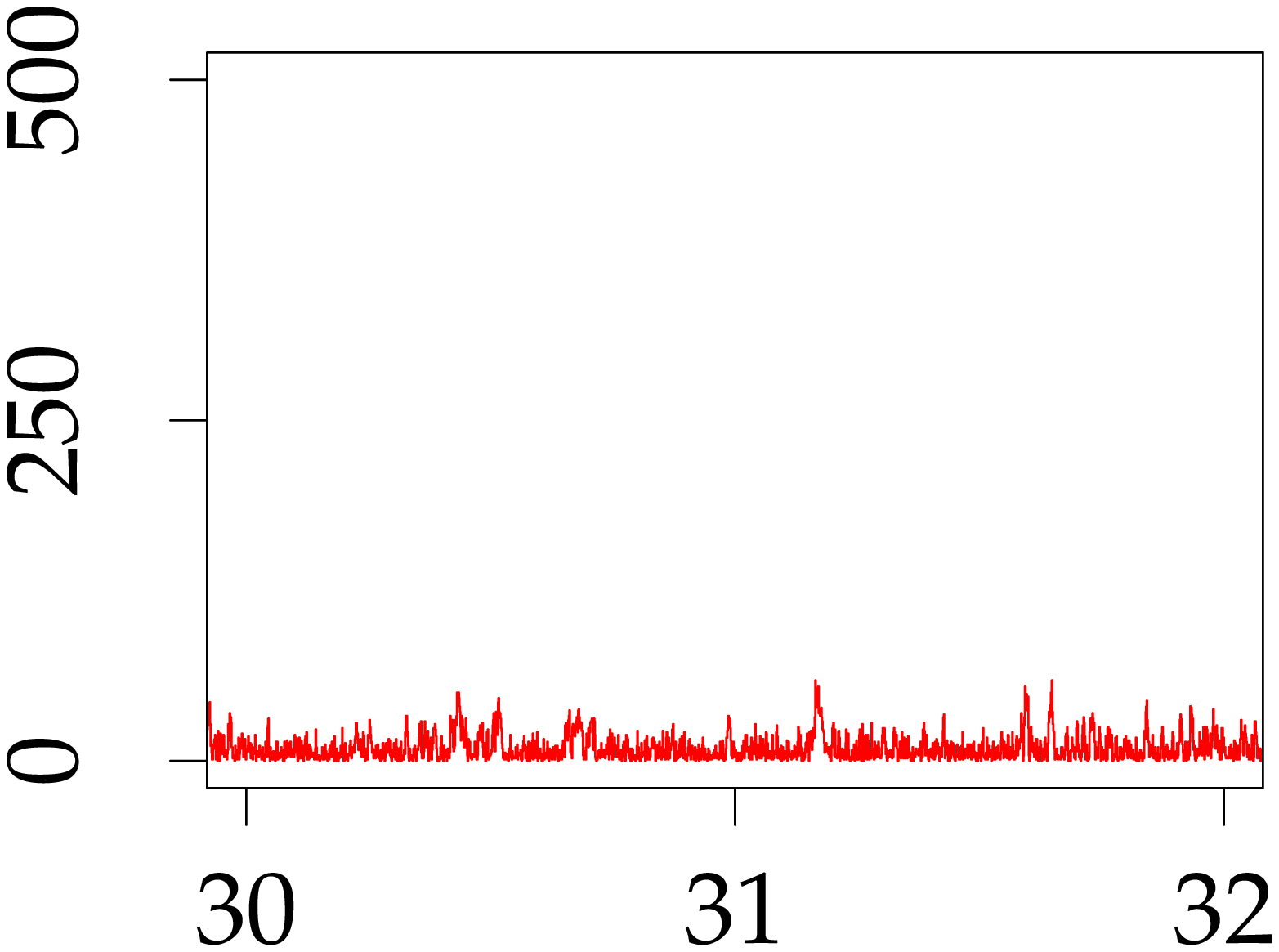} \\
 \multicolumn{2}{c}{\small{$a$ = 0.2}}\\ 
\includegraphics[height=\hite, width=0.45\textwidth]{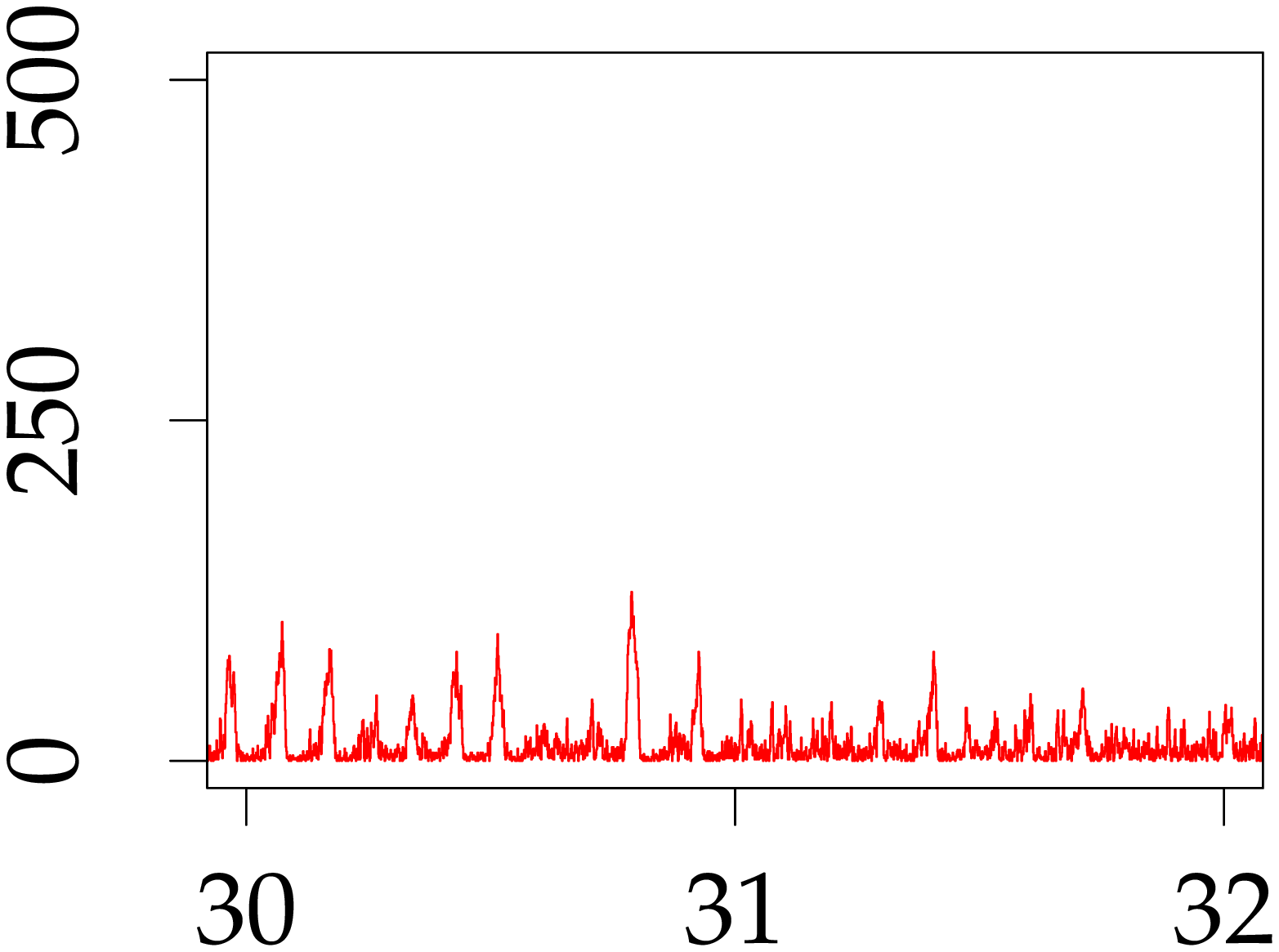} &
\includegraphics[height=\hite, width=0.45\textwidth]{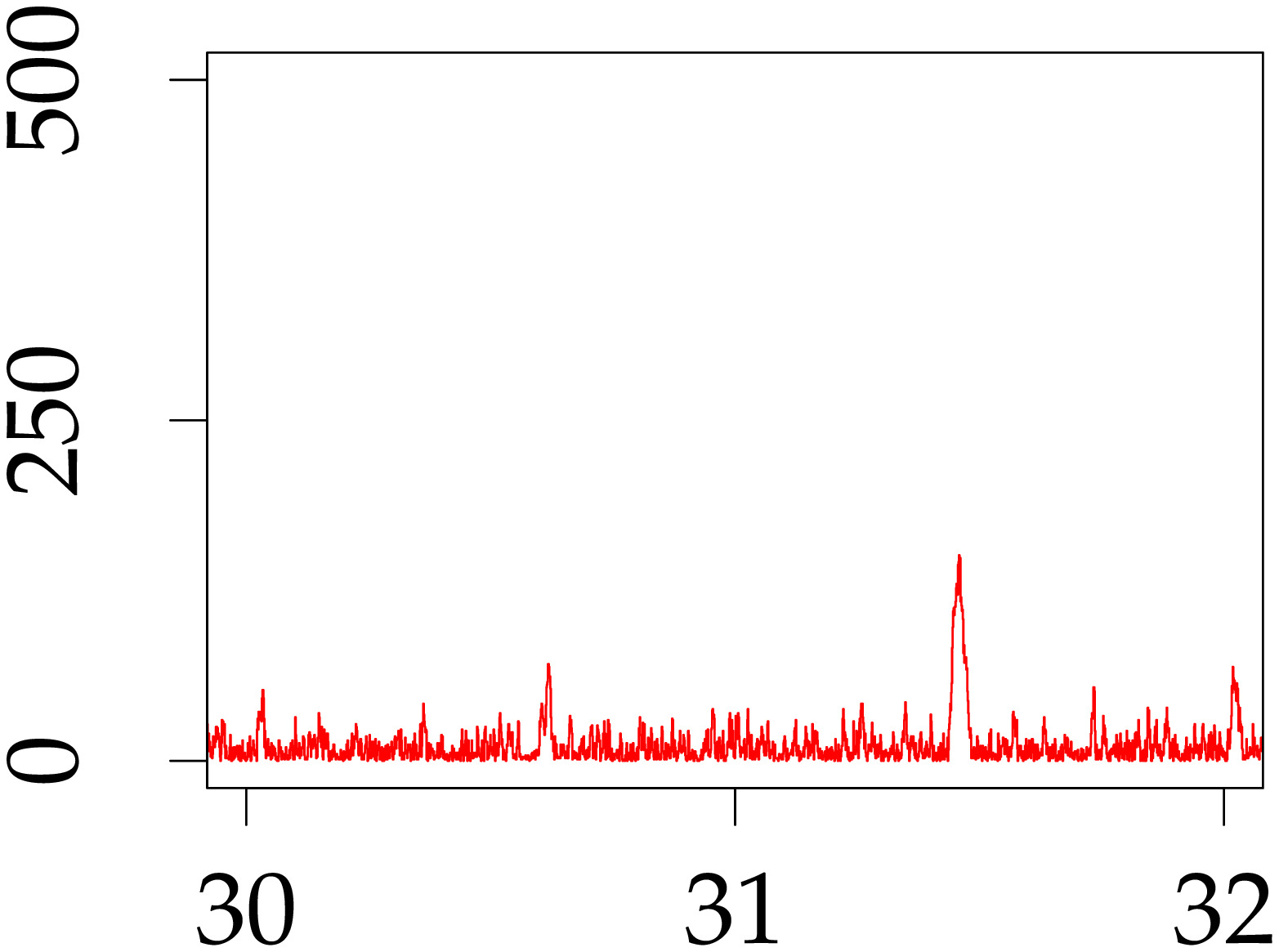} \\
 \multicolumn{2}{c}{\small{$a$ = 0.6}}\\ 
\includegraphics[height=\hite, width=0.45\textwidth]{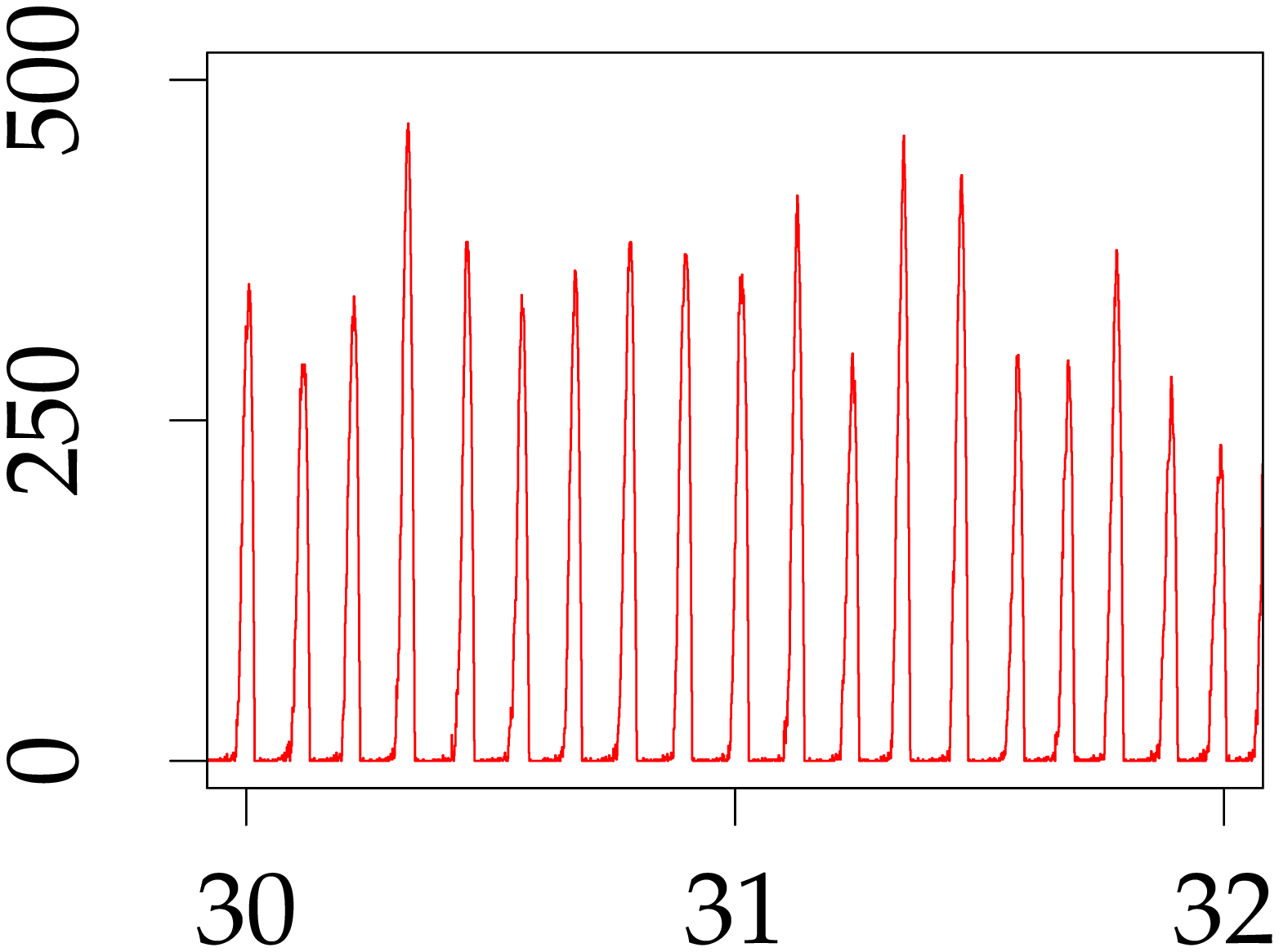} &
\includegraphics[height=\hite, width=0.45\textwidth]{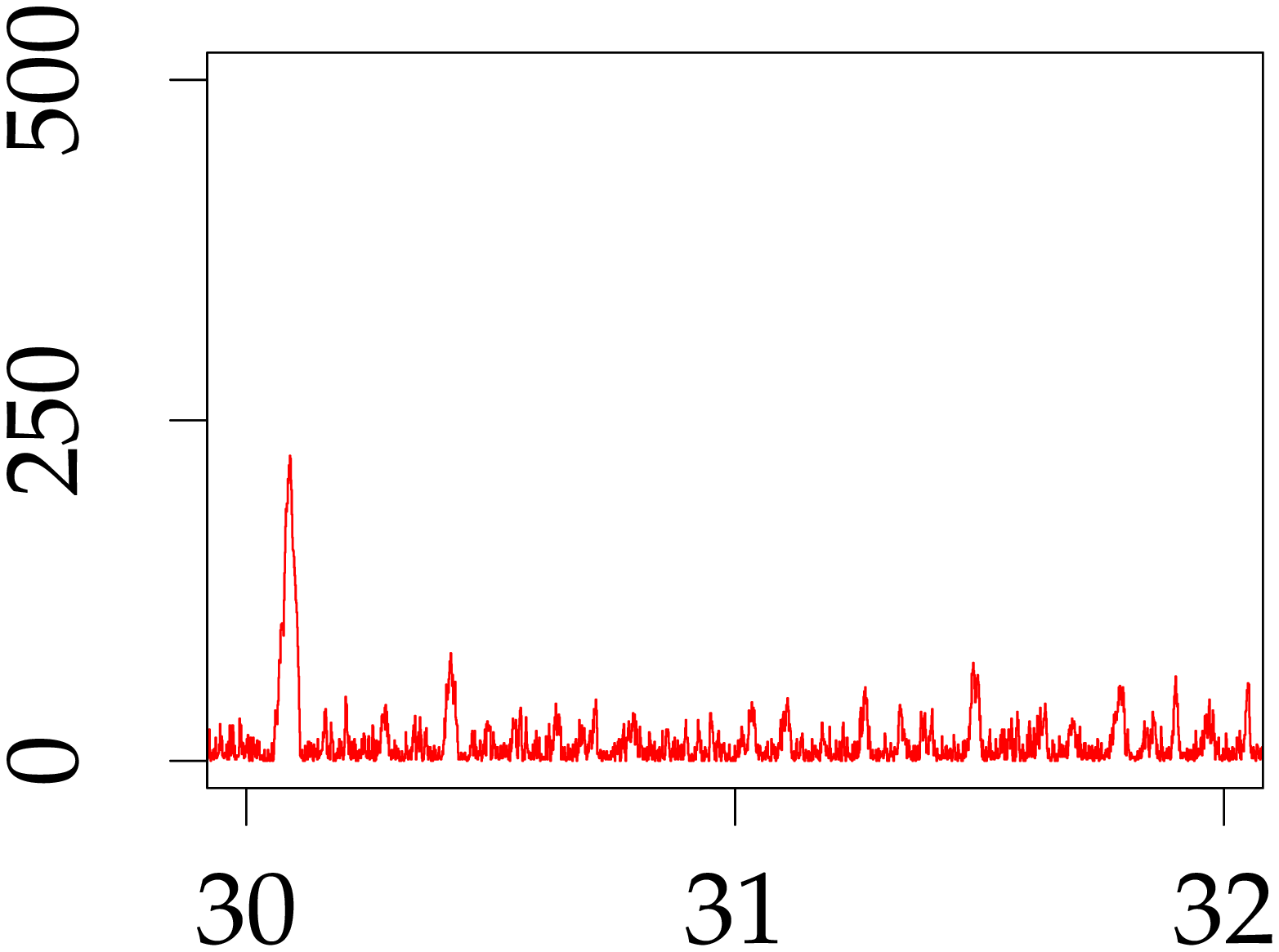} \\
 \multicolumn{2}{c}{\small{$a$ = 1.0}}\\ 
	\multicolumn{2}{c}{\small{Time (ms)}}\\
	\end{tabular}
\caption{\small{Traces from simulation of single bottleneck link with 100 RCP sources, $\tau_1= 10$ ms, $\tau_2= 50$ ms, $C=100$ packets per ms, and target link utilisation of 95\%.}}
%\end{center}
\label{fig:Qvsa}
\end{figure}

\section{Local Hopf Bifurcation Analysis}
\label{sec:rcp2dhba}
In stability analysis, we have shown that by excluding feedback based on queue size, we can increase the range of protocol parameter $a$, for which the stability is guaranteed. Also, we have established that the system loses local stability via a Hopf bifurcation and leads to the emergence of limit cycles. An important design objective of the system is not only to ensure the local stability for a wide range of parameters, but also to make sure that any loss of stability always result in stable limit cycles of small amplitude. So it is natural to study the characteristics of the bifurcating periodic solutions. To that end, we do a local Hopf bifurcation analysis.\\

% It enables us to develop a better understanding of the effect of the queue feedback on the nature and stability of the bifurcating limit cycles.
% Moreover, to have a better understanding of the non-trivial role played by the queue size feedback, it is not enough to consider only the linear approximation of the model. But also to consider the quadratic and cubic terms as well in the Taylor series.\\ 
The analytical framework employed to investigate the nature of the limit cycles are the Poincar{\'e} normal form and the center manifold theorem, which are outlined in the Appendix. The analysis relies on the linear, quadratic and cubic terms in the
Taylor series expansion of (\ref{eq:simplified_model}) and (\ref{eq:simplified_withoutQ_model}), about equilibrium, which have been tabulated in Table \ref{tab:tayexp}. The coefficients not listed in the table do not exist.
\begin{table}[b!]
\renewcommand{\arraystretch}{2.25}
\centering
\caption{Linear, quadratic and cubic terms of a Taylor series expansion of~(\ref{eq:simplified_model}) and ~(\ref{eq:simplified_withoutQ_model}) for RCP with small buffers. For brevity, the parameter $\tau_1,\tau_2$ represent the round-trip delays, $\xi_x$ represents $f_x\big|_{(x^*,y^*,z^*)}$, $\xi_y$ represents $f_y\big|_{(x^*,y^*,z^*)}$, $\xi_z$ represents $f_z\big|_{(x^*,y^*,z^*)}$, $\xi_{xx}$ represents $f_{xx}\big|_{(x^*,y^*,z^*)}$ and so on.}\label{tab:tayexp}
\begin{tabular}{ll}
\hline
$ \qquad \qquad \qquad \qquad \qquad \qquad b=0 $ & $ \qquad  \qquad b>0  $\\
\hline
%\vspace{-5mm}\\
$\xi_y=\xi_z \qquad \qquad \qquad \quad  \dfrac{-a}{(\tau_1 + \tau_2)} $  &  $   \dfrac{-a(1+(2R^*/C))}{(\tau_1 + \tau_2)}$\\
$\xi_{xy}=\xi_{xz} \qquad \qquad \qquad \dfrac{-a}{R^*(\tau_1 + \tau_2)}$ &$   \dfrac{-a(1+(2R^*/C))}{R^*(\tau_1 + \tau_2)}$\\
$2\xi_{yy}=\xi_{yz}=2\xi_{zz}  \qquad \;\; 0$& $   \dfrac{-2a}{(\tau_1+\tau_2)\sqrt{bCR^*}}$\\
$2\xi_{xyy}=\xi_{xyz}=2\xi_{xzz} \quad 0 $& $   \dfrac{-2a}{(\tau_1+\tau_2)R^*\sqrt{bCR^*}}$\\
$\xi_{yyz}=\xi_{yzz} \qquad \qquad \quad\; \;0 $&$  \dfrac{-3a}{(\tau_1+\tau_2)bCR^*}$\\
$\xi_{yyy}=\xi_{zzz} \qquad \qquad \quad\; \;0 $&$  \dfrac{-a}{(\tau_1+\tau_2)bCR^*}$\\
\hline
\hline
\end{tabular}
\\
%\vspace{-5mm}
\end{table}
%The Appendix contains the requisite analysis, and we present some computational results for RCP under consideration.
%{R^2 (\tau_1 + \tau_2)}
\subsection{Without queue feedback}
\begin{figure}[hbtp!]
\centering
\psfrag{x}{$\vartheta$}
\psfrag{y}[c]{$\tilde{f}(\vartheta)$}
\psfrag{0.000}{\small{\hspace{2.5mm}$0$}}
\psfrag{0.785}{\small{\hspace{1mm}$\pi/4$}}
\psfrag{1.570}{\small{\hspace{1mm}$\pi/2$}}
\psfrag{2.356}{\small{$3\pi/4$}}
\psfrag{3.141}{\small{\hspace{2.5mm}$\pi$}}
\psfrag{beta}{\small{$\beta_2$}}
\psfrag{myu2}{\small{$\mu_2$}}
\psfrag{105}{{\small{$\times 10^{-5}$}}}
\psfrag{107}{{\small{$\times 10^{-7}$}}}
\psfrag{kappa}{\hspace{-13mm}Bifurcation parameter,  $\kappa$}
\includegraphics[trim=0cm 0cm 0cm 1cm, clip=true,width=3.25in,height=2.25in]{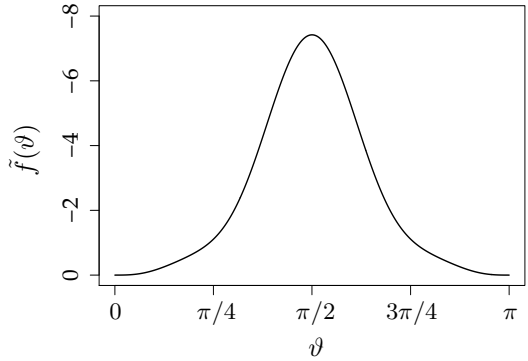}
\vspace{-2mm}
\caption{ The plot of $\tilde{f}(\vartheta)$ as $\vartheta$ varies. As $\tilde{f}(\vartheta)<0$ and hence $\mathbf{Re}\big(c_1(0)\big)<0$, the Hopf bifurcation is super-critical and the limit cycles are orbitally stable. }
%\vspace{-5mm}
\label{fig:realc10}
\end{figure}

In this section, we consider the small buffer model of RCP without queue feedback and perform the necessary calculations to determine the type of Hopf bifurcation and the asymptotic form of the bifurcation solutions as local instability just sets in. For now, we will only be concerned with the first Hopf bifurcation.
As outlined in the appendix, the stability and direction of the bifurcating limit cycles can be determined from the sign of first Lyapunov coefficient($\mu_2$) and Floquet exponent($\beta_2$),\\
where
\begin{eqnarray}
\mu_2 &&\hspace{-6mm}= \dfrac{-\operatorname{Re}[c_1(0)]}{\alpha'(0)}\quad \beta_2= 2\operatorname{Re}[c_1(0)].\nonumber
%\beta &&\hspace{-6mm}= \epsilon^2\beta_2+\mathcal{O}(\epsilon^4)\quad \beta_2 = 2\operatorname{Re}[c_1(0)]\quad \epsilon = \sqrt{\frac{\mu}{\mu_2}}\nonumber\\\label{49}
\end{eqnarray}
Using the definitions outlined in the appendix and the values
from Table \ref{tab:tayexp}, the expression for $\operatorname{Re}[c_1(0)]$ has been calculated as
% \begin{equation}
% \operatorname{Re}[c_1(0)] = \frac{2\pi~\tilde{f}(\omega_0\tau_1)}{
% {(\gamma C)}^2 (\tau_1 + \tau_2)}
% \label{eq:realc10}
% \end{equation}
% where
% \begin{equation}
% \begin{aligned}
% \tilde{f}(\omega_0\tau_1) &=-2\pi\sin^4(\omega_0\tau_1)-\pi\sin^2(\omega_0\tau_1)\cos^2(2\omega_0\tau_1)\\
% &-2\cos(2\omega_0\tau_1)\sin^3(\omega_0\tau_1)\\
% &-\cos(2\omega_0\tau_1)\sin^2(\omega_0\tau_1)\cos(\omega_0\tau_1)(\pi-2\omega_0\tau_1).\nonumber
% \end{aligned}
% \end{equation}
\begin{equation}
\mathop{\mathrm{sign}}\Big(\mathbf{Re}\big(c_1(0)\big)\Big) = \mathop{\mathrm{sign}}\Bigg(\frac{2\pi~\tilde{f}(\vartheta)}{
{(\gamma C)}^2 (\tau_1 + \tau_2)}\Bigg),
\label{eq:realc10}
\end{equation}
where
\begin{equation}
\begin{aligned}
\vartheta =&\  \omega_0\tau_1 = \dfrac{\pi\tau_1}{\left(\tau_1+\tau_2\right)} \in (0,\pi),\\
\tilde{f}(\vartheta) =&-2\pi\sin^4(\vartheta)-\pi\sin^2(\vartheta)\cos^2(2\vartheta)\\
&-2\cos(2\vartheta)\sin^3(\vartheta)\\
&-\cos(2\vartheta)\sin^2(\vartheta)\cos(\vartheta)(\pi-2\vartheta).\nonumber
\end{aligned}
\end{equation}
From~(\ref{eq:realc10}), we can say that the sign of $\operatorname{Re}[c_1(0)]$ depends on $\tilde{f}(\vartheta)$. From Fig.~\ref{fig:realc10}, it is clear that the $\tilde{f}(\vartheta)$ is negative for all $\vartheta \in (0,\pi)$. Therefore $\mathbf{Re}\big(c_1(0)\big)$ is negative for all $\vartheta \in (0,\pi)$. We have already shown that $\alpha'(0)>0$ for all $\vartheta \in (0,\pi)$. Hence, 
we have $\mu_2>0$ and $\beta_2<0$ which enables us to conclude that the system undergoes a super-critical Hopf and the limit cycles are asymptotically orbitally stable.
% \begin{figure}[h!]
% \centering
% \psfrag{x}{$\omega_0\tau_1$}
% \psfrag{y}{$\tilde{f}(\omega_0\tau_1)$}
% \psfrag{0.000}{\small{$0$}}
% \psfrag{0.785}{\small{$\pi/4$}}
% \psfrag{1.570}{\small{$\pi/2$}}
% \psfrag{2.356}{\small{$3\pi/4$}}
% \psfrag{3.141}{\small{$\pi$}}
% \psfrag{beta}{\small{$\beta_2$}}
% \psfrag{myu2}{\small{$\mu_2$}}
% \psfrag{105}{{\small{$\times 10^{-5}$}}}
% \psfrag{107}{{\small{$\times 10^{-7}$}}}
% \psfrag{kappa}{\hspace{-13mm}Bifurcation parameter,  $\kappa$}
% %\includegraphics[trim=0cm 0cm 0cm 1cm, clip=true,width=3.25in,height=2.25in]{./R_Plot/mubetacurve.eps}
% \includegraphics[trim=0cm 0cm 0cm 1cm, clip=true,width=3.25in,height=2.25in]{./R_Plot/realc10.eps}
% \vspace{-2mm}
% \caption{ The plot for $\tilde{f}(\omega_0\tau_1)$ for all possible values of $\omega_0\tau_1$. }
% %\vspace{-5mm}
% \label{fig:realc10}
% \end{figure}
% From Figure~\ref{fig:realc10}, it is clear that $\mathbf{Re}\big(c_1(0)\big)$ is negative for all $\vartheta \in (0,\pi)$. We have already shown that $\alpha'(0)>0$ for all $\vartheta \in (0,\pi)$. Hence, 
% we have $\mu_2>0$ and $\beta_2<0$ which enables us to conclude that the system undergoes a super-critical Hopf and the limit cycles are asymptotically orbitally stable.
\\

\subsection{With queue feedback}
\begin{figure}[hbtp!]
\centering
\psfrag{x}{$\vartheta$}
\psfrag{y}[c]{$\mu_2$}
\psfrag{0.000}{\small{\hspace{2.5mm}$0$}}
\psfrag{0.785}{\small{\hspace{1mm}$\pi/4$}}
\psfrag{1.570}{\small{\hspace{1mm}$\pi/2$}}
\psfrag{2.356}{\small{$3\pi/4$}}
\psfrag{3.141}{\small{\hspace{2.5mm}$\pi$}}
\psfrag{beta}{\small{$\beta_2$}}
\psfrag{myu2}{\small{$\mu_2$}}
\psfrag{105}{{\small{$\times 10^{-5}$}}}
\psfrag{107}{{\small{$\times 10^{-7}$}}}
\psfrag{kappa}{\hspace{-13mm}Bifurcation parameter,  $\kappa$}
\includegraphics[trim=0cm 0cm 0cm 1cm, clip=true,width=3.25in,height=2.25in]{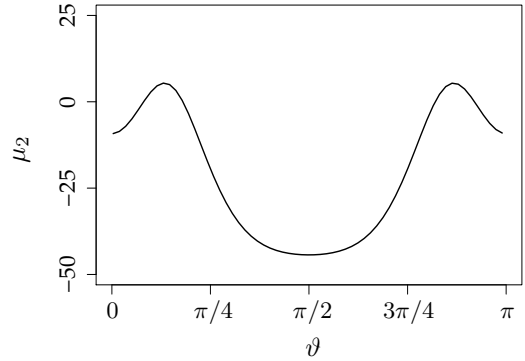}
\vspace{-2mm}
\caption{ Effect of $\vartheta$ on the criticality of the Hopf bifurcation: sub-critical if $\mu_2<0$ and super-critical if $\mu_2>0$.}
%\vspace{-5mm}
\label{fig:myu2Vsteta}
\end{figure}

\begin{figure}[hbtp!]
\centering
\psfrag{x}{$\rho^*$}
\psfrag{y}[c]{$\mu_2$}
\psfrag{0.000}{\small{\hspace{2.5mm}$0$}}
\psfrag{0.785}{\small{\hspace{1mm}$\pi/4$}}
\psfrag{1.570}{\small{\hspace{1mm}$\pi/2$}}
\psfrag{2.356}{\small{$3\pi/4$}}
\psfrag{3.141}{\small{\hspace{2.5mm}$\pi$}}
\psfrag{beta}{\small{$\beta_2$}}
\psfrag{myu2}{\small{$\mu_2$}}
\psfrag{105}{{\small{$\times 10^{-5}$}}}
\psfrag{107}{{\small{$\times 10^{-7}$}}}
\psfrag{kappa}{\hspace{-13mm}Bifurcation parameter,  $\kappa$}
\includegraphics[trim=0cm 0cm 0cm 1cm, clip=true,width=3.25in,height=2.25in]{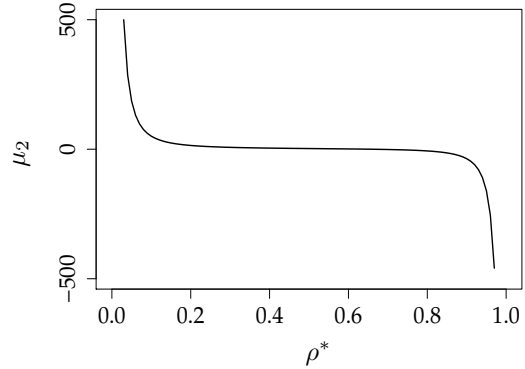}
\vspace{-2mm}
\caption{ The plot of $\mu_2$ as equilibrium utilization ($\rho^*$) varies. For low $\rho^*$, $\mu_2>0$ and hence the Hopf bifurcation is super-critical. For high $\rho^*$, $\mu_2<0$ which implies that the Hopf bifurcation is sub-critical.   }
%\vspace{-5mm}
\label{fig:myu2Vsrho}
\end{figure}

We now consider the RCP model which uses both rate mismatch and queue size feedback. Let us denote the equilibrium utilisation as 
\begin{equation}
 \rho^*=\frac{2R^*}{C}
\end{equation}
%Let $C$\ =\ 2, then the equilibrium utilisation is same as equilibrium rate i.e.\ $\rho^*=R^*$.
Using the calculations outlined in the Appendix, we obtain the closed-form analytical expression for $\mu_2$ as
\begin{equation}
\mu_2 = \frac{ \mathbf{Re} (\tilde{g}(\vartheta,\rho^*) \bar{D})}{\mathbf{Re}(i(1+\rho^*)\bar{D})}
\end{equation}
where
\begin{equation}
\label{eq:myu2_withq}
\begin{aligned}
\tilde{g}(\vartheta,\rho^*) =&\dfrac{\sin(\vartheta)}{\rho^*(1-\rho^*)} \left(2i\sin(\vartheta)- \dfrac{2\cos(2\vartheta)+i\sin(\vartheta)}{\cos(2\vartheta)+2i\sin(\vartheta)} \right) \\
&+\frac{(1+\rho^*)(-\sin(\vartheta)+i\cos(2\vartheta))}{(\rho^*)^2(\cos(2\vartheta)+2i\sin(\vartheta))} \\
&+\frac{2sin^2(\vartheta)(-4\sin(\vartheta)+3i\cos(2\vartheta))}{(1-\rho^*)^2(1+\rho^*)(\cos(2\vartheta)+2i\sin(\vartheta))} \\
&-\frac{3i(3\sin(\vartheta)-\sin(3\vartheta))}{4(1-\rho^*)^2\sin(\vartheta)}
\end{aligned}
\end{equation}
%&\frac{(1+\rho^*)(-\sin(\vartheta)+i\cos(2\vartheta))}{(\rho^*)^2(\cos(2\vartheta)+2i\sin(\vartheta))} + \frac{\sin^2(\vartheta)}{\rho^*(1-\rho^*)} \\
%&+ 2i\\ + \frac{\sin^2(\vartheta)}{\rho^*(1-\rho^*)}
It is to be noted that the protocol parameter $a$ has no effect on the nature of the Hopf bifurcation.
The criticality of the Hopf bifurcation and the oscillation amplitude depends on the value of $\mu_2$.
In this case, we go for numerical examples to study the nature of Hopf bifurcation.\\ \\
\textit{Numerical Example 1}: Let us consider $b=0.022$ which corresponds to equilibrium utilisation of $\rho^*=0.9$.  The values of $\mu_2$ computed using~\eqref{eq:myu2_withq} for all $\vartheta \in (0,\pi)$ has been plotted in Fig.~\ref{fig:myu2Vsteta}. From Fig.~\ref{fig:myu2Vsteta}, we can observe that both super-critical and sub-critical Hopf occurs, and the criticality varies with $\vartheta$ which is a function of $\tau_1$ and $\tau_2$.\\ \\
\textit{Numerical Example 2}: Now we assume $\tau_2=2\tau_1$ which corresponds to $\vartheta=\pi/3$, and vary the equilibrium utilisation to analyze its impact on the criticality of the Hopf bifurcation. Fig.~\ref{fig:myu2Vsrho} shows the computed values of $\mu_2$ for $\rho^* \in (0,1)$. As the equilibrium utilisation increases, the criticality of the bifurcation changes from super-critical to sub-critical and so there is the potential for large amplitude limit cycles which is undesirable.\\ \\
To summarise, the results of the local Hopf bifurcation analysis enables us to say that the RCP without queue feedback undergoes super-critical Hopf bifurcation and leads to orbitally stable limit cycles. Whereas,
we observed sub-critical Hopf bifurcation for some parameter values if the queue size feedback is incorporated into the RCP model. The super-critical case is highly desirable from a performance viewpoint than the sub-critical one, since the sub-critical Hopf bifurcation results in sharp loss of system stability. Therefore, the removal of queue size feedback would be the most appropriate design choice for RCP.\\


\subsection{Numerical computations}

In a scalar non-linear equation with two discrete delays, both the delays play an important role in determining the type of the Hopf bifurcation. As we have just witnessed that even with the same non-linear term, by simply changing the values of the delay, we can change the type of the Hopf bifurcation.

We now validate the analytical results using some numerical examples.
\subsubsection{When round-trip times vary}
\begin{figure}[h!]
\centering
\psfrag{R}[][][2]{Amplitude of oscillation}
\psfrag{kappa}[][][2]{Bifurcation parameter, $\kappa$}
\psfrag{0.95}[][][2]{$0.95$}
\psfrag{1.00}[][][2]{\scriptsize $1.00$}
\psfrag{1.01}[][][2]{\scriptsize $1.01$}
\psfrag{1.02}[][][2]{\scriptsize $1.02$}
\psfrag{1.03}[][][2]{\scriptsize $1.03$}
\psfrag{1.04}[][][2]{\scriptsize $1.04$}
\psfrag{1.05}[][][2]{\scriptsize $1.05$}
\psfrag{0}[][][2]{\scriptsize $0$}
\psfrag{10}[][][2]{\scriptsize $10$}
\psfrag{20}[][][2]{\scriptsize $20$}
\includegraphics[scale = 0.9,trim=0cm 0cm 0cm 1.7cm, clip=true,width=3.25in,height=2.25in]{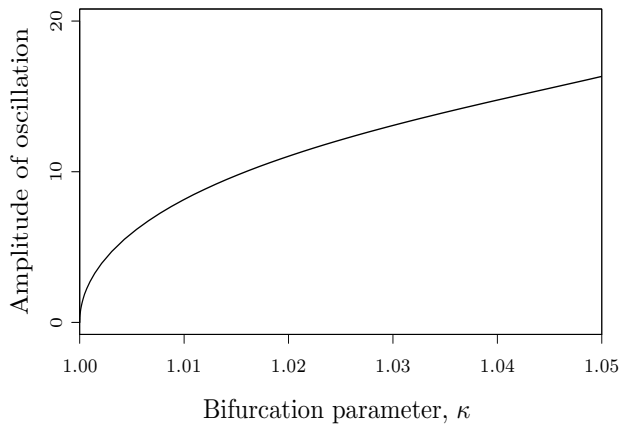}
\vspace{-2mm}
\caption{Bifurcation diagram highlighting that the system undergoes a super-critical Hopf bifurcation at $\kappa=1$. The parameter values used are  $a=2.16$, $b=0.022$, $C=100$, $\tau_1=10$ and $\tau_2=70$ ($\vartheta=\pi/8$).}
%\vspace{-5mm}
\label{fig:rcp2d_withq_bfd_wrt_tau_super}
\vspace{-2mm}
\end{figure}

\begin{figure}[h!]
\centering
\psfrag{R}[][][2]{ Amplitude of oscillation}
\psfrag{kappa}[][][2]{ Bifurcation parameter, $\kappa$}
\psfrag{0.95}[][][2]{\scriptsize $0.95$}
\psfrag{0.96}[][][2]{\scriptsize $0.96$}
\psfrag{0.97}[][][2]{\scriptsize $0.97$}
\psfrag{0.98}[][][2]{\scriptsize $0.98$}
\psfrag{0.99}[][][2]{\scriptsize $0.99$}
\psfrag{1.00}[][][2]{\scriptsize $1.00$}
\psfrag{1.01}[][][2]{\scriptsize $1.01$}
\psfrag{1.02}[][][2]{\scriptsize $1.02$}
\psfrag{1.03}[][][2]{\scriptsize $1.03$}
\psfrag{1.04}[][][2]{\scriptsize $1.04$}
\psfrag{1.05}[][][2]{\scriptsize $1.05$}
\psfrag{0}[][][2]{\scriptsize $0$}
\psfrag{10}[][][2]{\scriptsize $10$}
\psfrag{20}[][][2]{$20$}
\psfrag{30}[][][2]{$30$}
\psfrag{40}[][][2]{$40$}
\psfrag{4}[][][2]{\scriptsize $4$}
\psfrag{2}[][][2]{\scriptsize$2$}
\psfrag{5}[][][2]{\scriptsize $5$}
\psfrag{60}[][][2]{$60$}
\psfrag{104}{\small{$\times 10^{-4}$}}
\includegraphics[scale = 0.9,trim=0cm 0cm 0cm 1.7cm, clip=true,width=3.25in,height=2.25in]{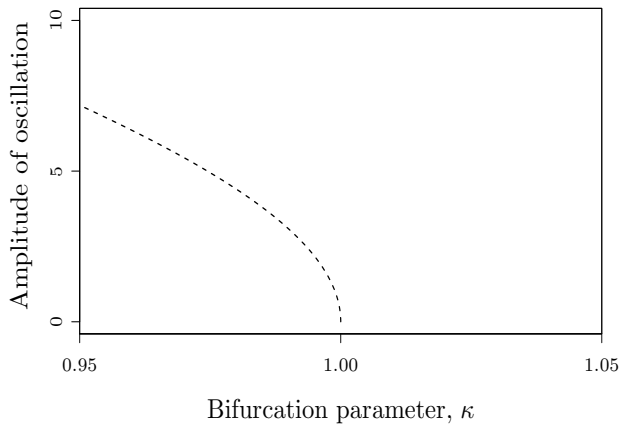}
\vspace{-2mm}
\caption{Bifurcation diagram showing the existence of a sub-critical Hopf for $a=0.87$, $b=0.022$, $C=100$, $\tau_1=10$ and $\tau_2=15$ ($\vartheta=2\pi/5$).}
%\vspace{-5mm}
\label{fig:rcp2d_withq_bfd_wrt_tau_sub}
\end{figure}

$\textit{Numerical Example 1 (Super-critical)}$: Let us consider the RCP system with $C=100$, $\tau_1=10$ and $\tau_2=70$ which corresponds to $\vartheta = \pi/8$. We set $b=0.022$ which corresponds to equilibrium utilization of 90\% of link capacity i.e., $\rho^*=0.9$. For these values, using \eqref{eq:nsstability_relation}, we obtain $a=2.16$, for $\kappa_c =1$. From Figure \ref{fig:myu2Vsteta}, we can observe that the value of  $ \mu_2$ is positive, implying that the system undergoes a super-critical Hopf bifurcation. The bifurcation diagram drawn using the Matlab package DDE-Biftool \citep{ddebiftool1,ddebiftool2} is shown in Figure \ref{fig:rcp2d_withq_bfd_wrt_tau_super}. As expected, it shows that the system loses local stability via a super-critical Hopf bifurcation, as the bifurcation parameter crosses the critical threshold ($\kappa_c =1$). To validate this, numerical simulations obtained using XPPAUT \citep{xppaut2002} are shown in Figure \ref{fig:rcp2d_withq_ns_wrt_tau_super}. For $\kappa=0.95$, the system converges to the equilibrium rate, $R^*=45$ (see Figure \ref{fig:rcp2d_withq_ns_wrt_tau_super}(a)). Whereas, for $\kappa = 1.05 > \kappa_c$ i.e. after the bifurcation, the system leads to the emergence of stable limit cycles.\\
%%%%%%%%%%%%%%%%%%%%%%%%%%%%
\newcommand{\supwdth}{0.5\textwidth}
\begin{figure}[h!]
\psfrag{time}{\hspace{-0.1cm}  \scriptsize Time}
\psfrag{Rate}{\hspace{-0.7cm} Rate, $R(t)$}
\psfrag{0}{\footnotesize{$0$}}
\psfrag{5}{\hspace{-0.05cm}\scriptsize{$5$}}
\psfrag{30}{\hspace{-0.1cm}\scriptsize{$30$}}
\psfrag{15}{\hspace{-0.1cm}\scriptsize{$15$}}
\psfrag{60}{\hspace{-0.1cm}\scriptsize{$60$}}
\psfrag{25000}{\hspace{-0.1cm}\scriptsize{$25000$}}
\psfrag{5000}{\hspace{-0.1cm}\scriptsize{$5000$}}
\psfrag{10000}{\hspace{-0.1cm}\scriptsize{$10000$}}
\psfrag{20000}{\hspace{-0.1cm}\scriptsize{$20000$}}
\psfrag{0.5}{\hspace{-0.1cm}\scriptsize{$0.95$}}
\psfrag{1.0}{\hspace{-0.1cm}\scriptsize{$1.0$}}
\psfrag{1.5}{\hspace{-0.1cm}\scriptsize{$1.05$}}
\psfrag{3}{\hspace{-0.1cm}\scriptsize{$3$}}
\psfrag{1}{\hspace{-0.1cm}\scriptsize{$1$}}
\psfrag{2}{\hspace{-0.1cm}\scriptsize{$2$}}
\begin{tabular}{c}
\subfloat[ $\kappa = 0.95$]{\includegraphics[width=\supwdth]{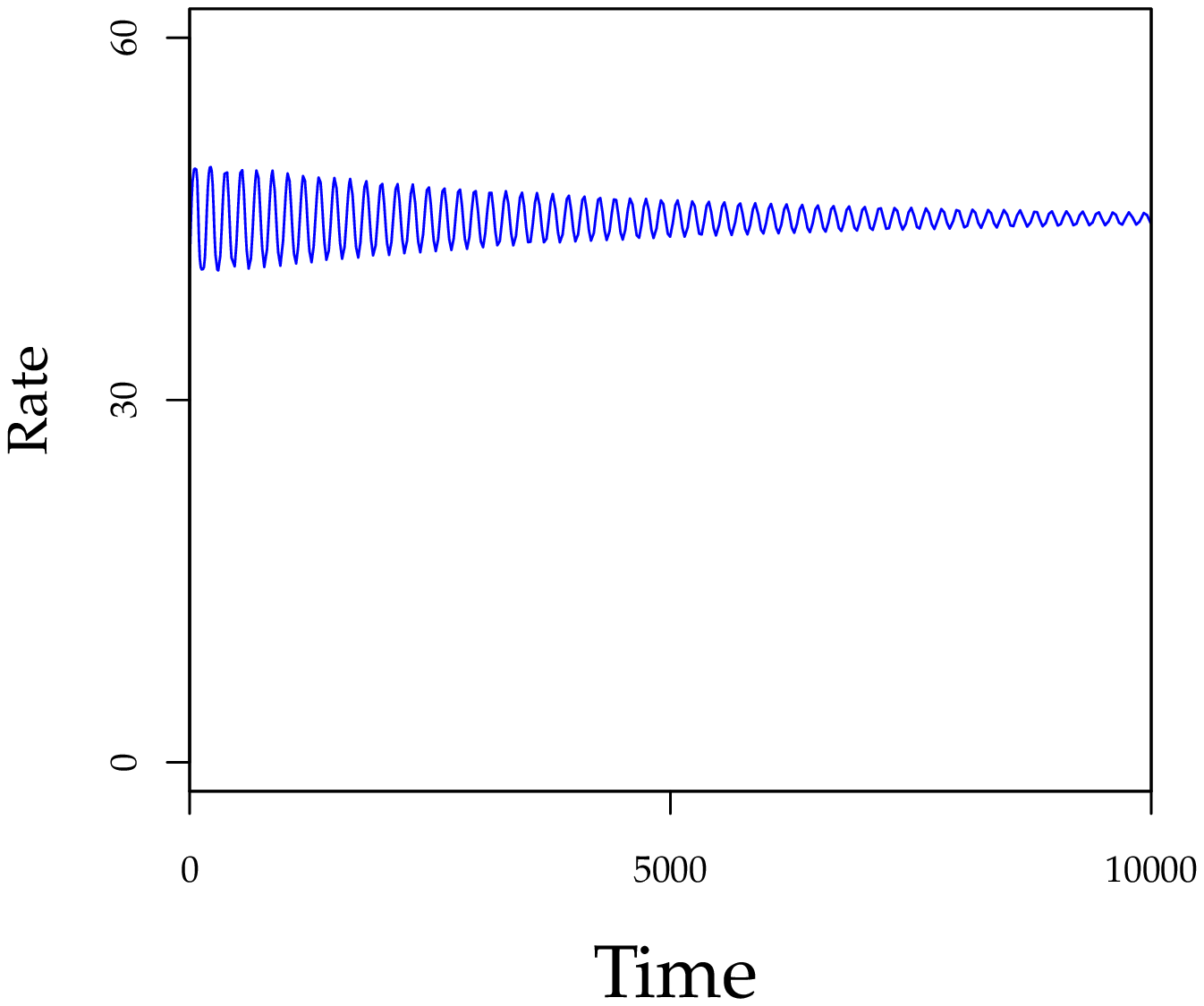}}\hspace{-5mm}
%\subfloat[ $\kappa = 0.97, R_0=2$]{\includegraphics[width=\supwdth]{tompecssuper2.eps}}\hspace{-5mm}
\subfloat[ $\kappa = 1.05$]{\includegraphics[width=\supwdth]{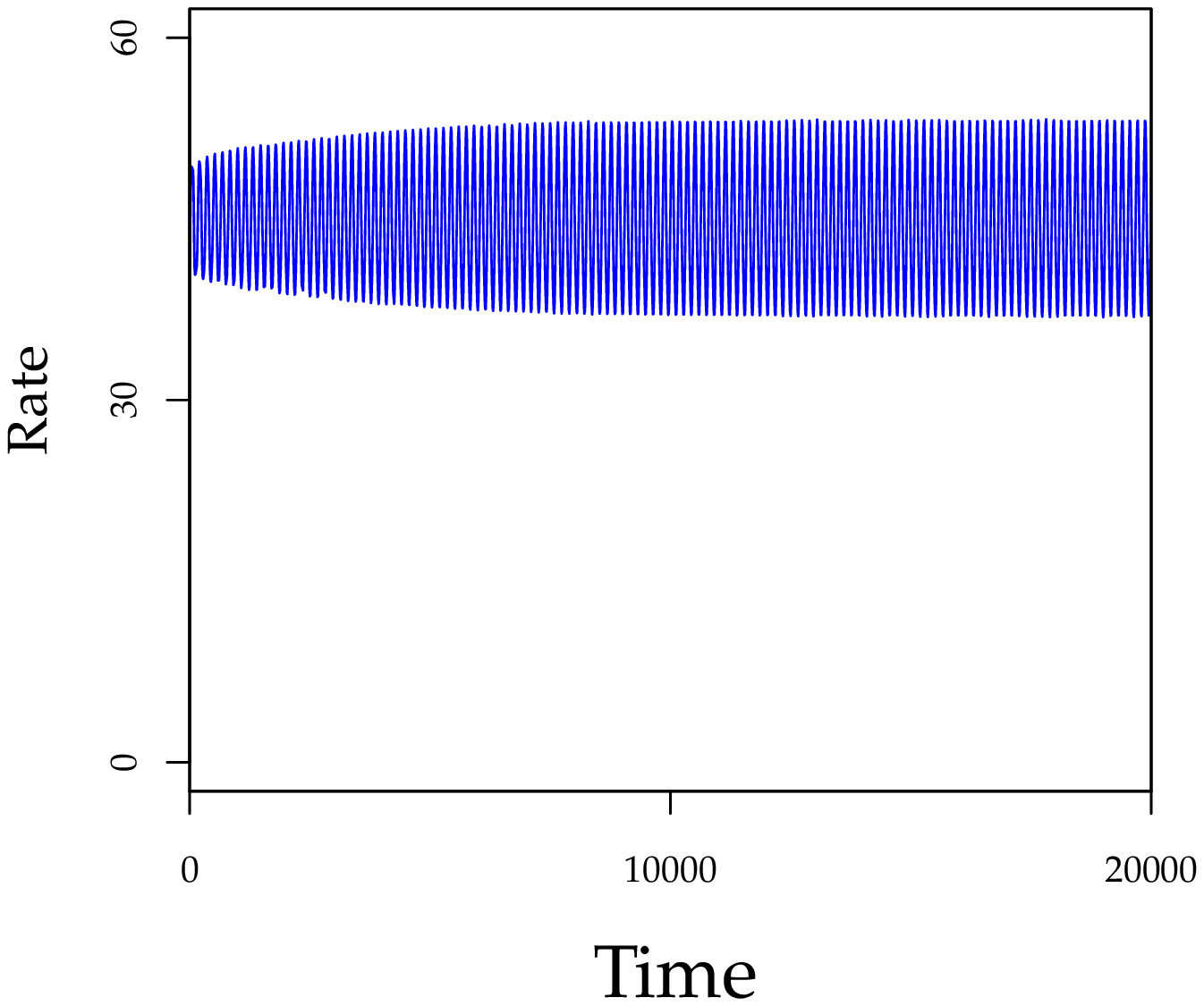}}
\end{tabular}
\caption{Numerical simulations illustrating that the system exhibits a super-critical Hopf bifurcation, as $\kappa$ increases  beyond the critical value. Time series are shown for the cases $\kappa < 1$ and $\kappa > 1$. The parameter values chosen are $a=2.16$, $b=0.022$, $C = 100$, $\tau_1 = 10$, $\tau_2=70$.} \label{fig:rcp2d_withq_ns_wrt_tau_super}
\end{figure}
\\
%%%%%%%%%%%%%%%%%%%%%%%%%%%%%%%%%%%%%%%%%%%%%%%%%%%%%
\\
$\textit{Numerical Example 2 (Sub-critical)}$: Consider $a=0.87$, $b=0.022$, $C=100$, $\tau_1=10$ and $\tau_1=15$ ($\vartheta=2\pi/5$). The system undergoes a Hopf bifurcation at $\kappa$ = 1. For these values, from Figure \ref{fig:myu2Vsteta}, we get $\mu_2 < 0$, implying that the system undergoes a sub-critical Hopf. We can also observe from Figure \ref{fig:rcp2d_withq_bfd_wrt_tau_sub} that there exists no stable limit cycle in the neighborhood, as the bifurcation parameter is varied beyond the critical threshold. Hence the Hopf bifurcation is sub-critical and the bifurcating limit cycles are unstable.
To illustrate the occurrence of a sub-critical Hopf, we present some numerical simulations in Figure \ref{fig:rcp2d_withq_ns_wrt_tau_sub}. For $\kappa=0.95$, the system converges to the stable equilibrium, $R^*=45$ (see Figure \ref{fig:rcp2d_withq_ns_wrt_tau_sub}(a)). Whereas, after the bifurcation i.e. for $\kappa > \kappa_c$, the previously stable fixed point now becomes unstable and also the solution would eventually jump to infinity (Figure \ref{fig:rcp2d_withq_ns_wrt_tau_sub}(b)).
\newcommand{\subwidth}{0.5\textwidth}
\begin{figure}[hbtp!]
\psfrag{time}{\hspace{-0.1cm}  \scriptsize Time}
\psfrag{Rate}{\hspace{-0.7cm}\vspace{2cm} Rate, $R(t)$}
\psfrag{0}{\hspace{-0.04cm}\scriptsize{$0$}}
\psfrag{5}{\hspace{-0.05cm}\scriptsize{$5$}}
\psfrag{10}{\hspace{-0.1cm}\scriptsize{$10$}}
\psfrag{15}{\hspace{-0.1cm}\scriptsize{$15$}}
\psfrag{30}{\hspace{-0.1cm}\scriptsize{$30$}}
\psfrag{25}{\hspace{0cm}\scriptsize{$25$}}
\psfrag{50}{\hspace{-0.1cm}\scriptsize{$50$}}
\psfrag{100}{\hspace{-0.1cm}\scriptsize{$100$}}
\psfrag{200}{\hspace{-0.1cm}\scriptsize{$200$}}
\psfrag{0.5}{\hspace{-0.1cm}\scriptsize{$0.95$}}
\psfrag{1.0}{\hspace{-0.1cm}\scriptsize{$1.0$}}
\psfrag{1.5}{\hspace{-0.1cm}\scriptsize{$1.05$}}
\psfrag{150}{\hspace{-0.1cm}\scriptsize{$150$}}
\psfrag{300}{\hspace{-0.1cm}\scriptsize{$300$}}
\psfrag{400}{\hspace{-0.1cm}\scriptsize{$400$}}
\psfrag{-800}{\hspace{-0.1cm}\scriptsize{$-800$}}
\psfrag{-400}{\hspace{-0.1cm}\scriptsize{$-400$}}
\psfrag{-300}{\hspace{-0.1cm}\scriptsize{$-300$}}
\psfrag{-600}{\hspace{-0.1cm}\scriptsize{$-600$}}
\psfrag{125}{\hspace{-0.1cm}\scriptsize{$125$}}
\psfrag{250}{\hspace{-0.1cm}\scriptsize{$250$}}
\psfrag{60}{\hspace{-0.1cm}\scriptsize{$60$}}
\psfrag{120}{\hspace{-0.1cm}\scriptsize{$120$}}
\psfrag{2500}{\hspace{-0.1cm}\scriptsize{$2500$}}
\psfrag{500}{\hspace{-0.1cm}\scriptsize{$500$}}
\psfrag{1000}{\hspace{-0.1cm}\scriptsize{$1000$}}
\psfrag{20000}{\hspace{-0.1cm}\scriptsize{$20000$}}
\psfrag{5000}{\hspace{-0.1cm}\scriptsize{$5000$}}
\begin{tabular}{c}
\subfloat[ $\kappa = 0.95$]{\includegraphics[width=\subwidth]{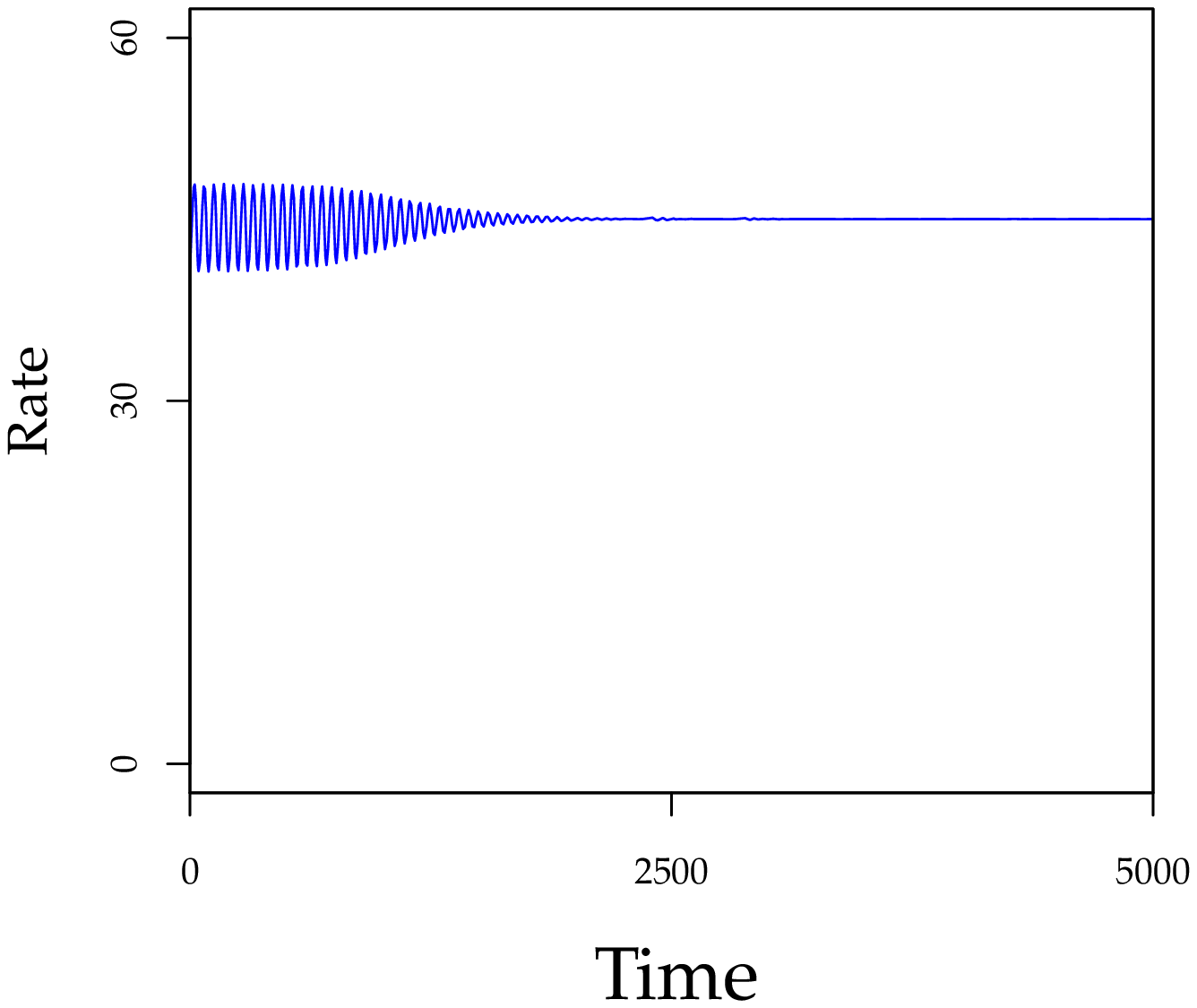}}\hspace{-5mm}
% \subfloat[ $\kappa = 0.97, R_0=2$]{\includegraphics[width=\subwdth]{tompecssub2.eps}}\hspace{-5mm}
\subfloat[ $\kappa = 1.05$]{\includegraphics[width=\subwidth]{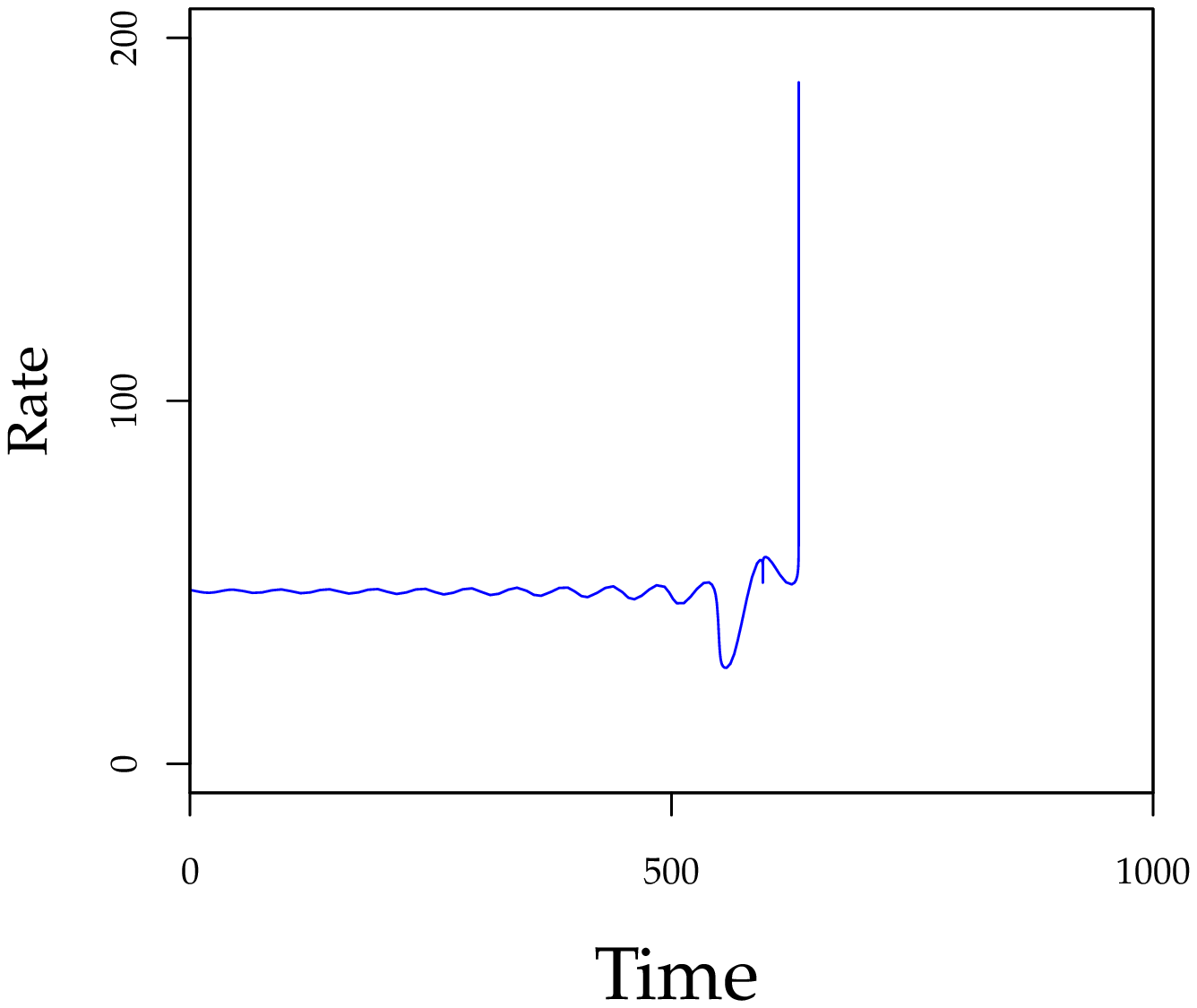}}
\end{tabular}
\caption{Numerical simulations highlighting that the system undergoes a sub-critical Hopf for the parameter values $\kappa_c= 1$, $a=0.827$, $\tau=100$, $C=10$ and $b=0.022$.} \label{fig:rcp2d_withq_ns_wrt_tau_sub}
\end{figure}
%%%%%%%%%%%%%%%%%%%% critc Vs util %%%%%%%%%%%%%%%%%%%%%%%%%%%%%%%%%%
\subsubsection{When target link utilization varies}
\begin{figure}[h!]
\centering
\psfrag{R}[][][2]{Amplitude of oscillation}
\psfrag{kappa}[][][2]{Bifurcation parameter, $\kappa$}
\psfrag{0.95}[][][2]{$0.95$}
\psfrag{1.00}[][][2]{\scriptsize $1.00$}
\psfrag{1.01}[][][2]{\scriptsize $1.01$}
\psfrag{1.02}[][][2]{\scriptsize $1.02$}
\psfrag{1.03}[][][2]{\scriptsize $1.03$}
\psfrag{1.04}[][][2]{\scriptsize $1.04$}
\psfrag{1.05}[][][2]{\scriptsize $1.05$}
\psfrag{0}[][][2]{\scriptsize $0$}
\psfrag{10}[][][2]{\scriptsize $10$}
\psfrag{20}[][][2]{\scriptsize $20$}
\psfrag{40}[][][2]{\scriptsize $40$}
\includegraphics[scale = 0.9,trim=0cm 0cm 0cm 1.7cm, clip=true,width=3.25in,height=2.25in]{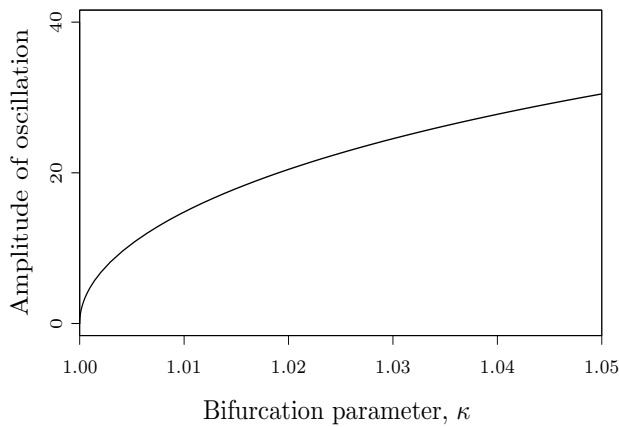}
\vspace{-2mm}
\caption{Bifurcation diagram highlighting that the system undergoes a super-critical Hopf bifurcation at $\kappa=1$. The parameter values used are  $a=1.17$, $b=0.736$, $C=100$, $\tau_1=10$ and $\tau_2=20$ ($\vartheta=\pi/3$).}
%\vspace{-5mm}
\label{fig:rcp2d_bfd_myuvsutil_sup}
\vspace{-2mm}
\end{figure}

\begin{figure}[h!]
\psfrag{time}{\hspace{-0.1cm}  \scriptsize Time}
\psfrag{Rate}{\hspace{-0.7cm} Rate, $R(t)$}
\psfrag{0}{\footnotesize{$0$}}
\psfrag{5}{\hspace{-0.05cm}\scriptsize{$5$}}
\psfrag{30}{\hspace{-0.1cm}\scriptsize{$30$}}
\psfrag{15}{\hspace{-0.1cm}\scriptsize{$15$}}
\psfrag{60}{\hspace{-0.1cm}\scriptsize{$60$}}
\psfrag{25000}{\hspace{-0.1cm}\scriptsize{$25000$}}
\psfrag{5000}{\hspace{-0.1cm}\scriptsize{$5000$}}
\psfrag{1000}{\hspace{-0.1cm}\scriptsize{$1000$}}
\psfrag{2000}{\hspace{-0.1cm}\scriptsize{$2000$}}
\psfrag{0.5}{\hspace{-0.1cm}\scriptsize{$0.95$}}
\psfrag{1.0}{\hspace{-0.1cm}\scriptsize{$1.0$}}
\psfrag{1.5}{\hspace{-0.1cm}\scriptsize{$1.05$}}
\psfrag{25}{\hspace{-0.1cm}\scriptsize{$25$}}
\psfrag{50}{\hspace{-0.1cm}\scriptsize{$50$}}
\psfrag{2500}{\hspace{-0.1cm}\scriptsize{$2500$}}
\begin{tabular}{c}
\subfloat[ $\kappa = 0.95$]{\includegraphics[width=\supwdth]{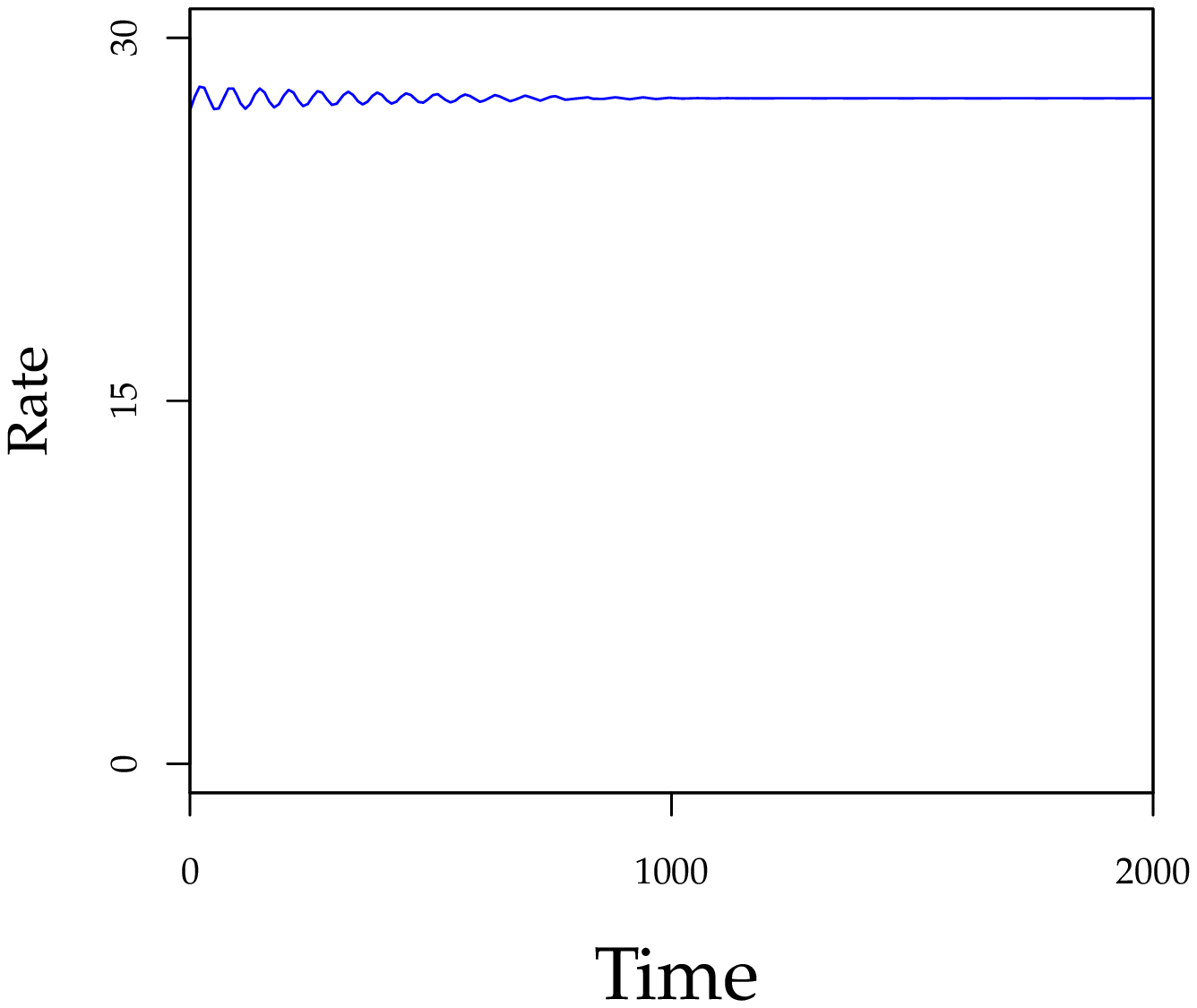}}\hspace{-5mm}
%\subfloat[ $\kappa = 0.97, R_0=2$]{\includegraphics[width=\supwdth]{tompecssuper2.eps}}\hspace{-5mm}
\subfloat[ $\kappa = 1.05$]{\includegraphics[width=\supwdth]{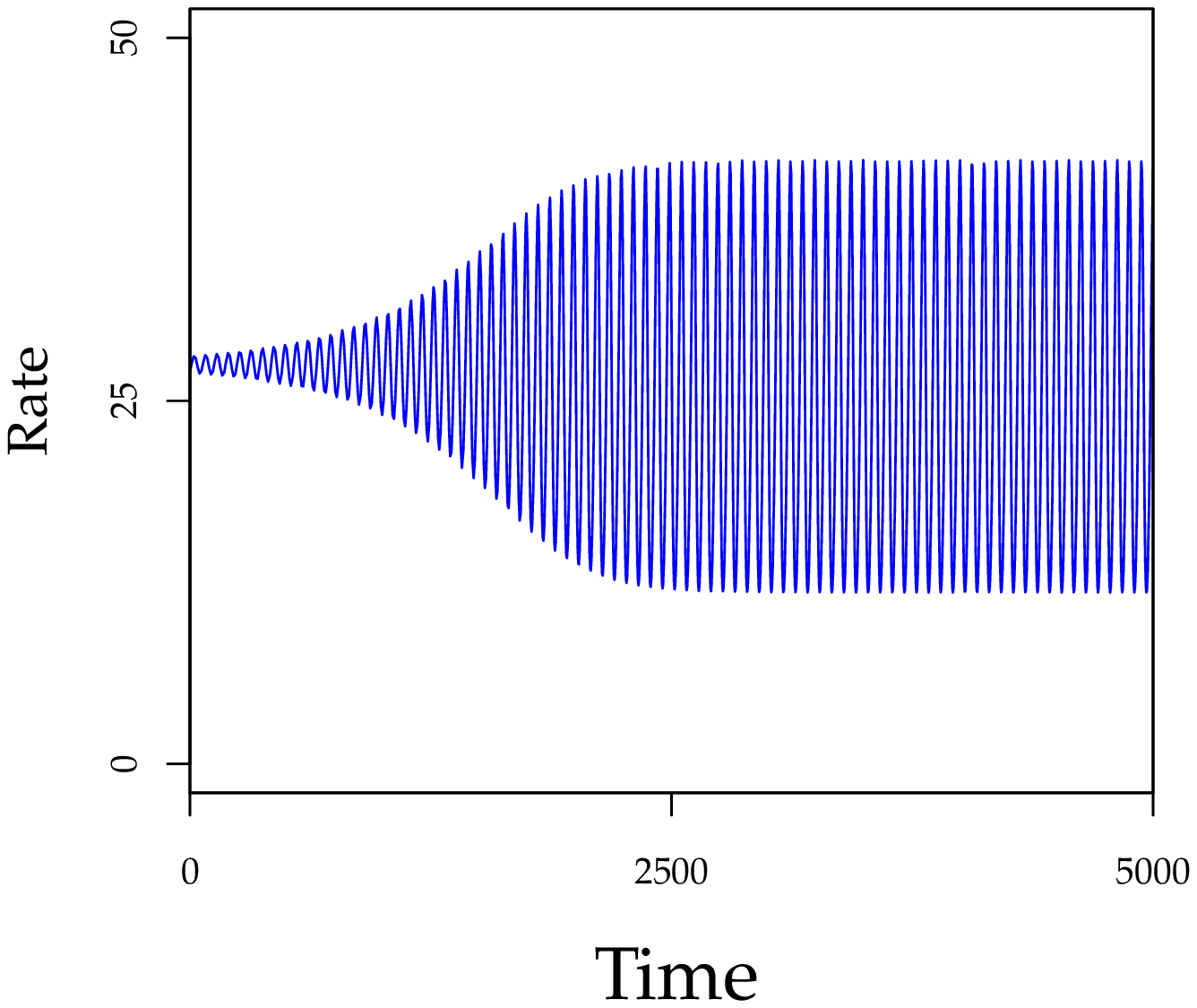}}
\end{tabular}
\caption{Numerical simulations illustrating that the system exhibits a super-critical Hopf bifurcation, as $\kappa$ increases  beyond the critical value. Time series are shown for the cases $\kappa < 1$ and $\kappa > 1$. The parameter values chosen are $a=1.17$, $b=0.736$, $C = 100$, $\tau_1 = 10$, $\tau_2=20$.} \label{fig:rcp2d_withq_ns_wrt_util_super}
\end{figure}

\begin{figure}[h!]
\centering
\psfrag{R}[][][2]{ Amplitude of oscillation}
\psfrag{kappa}[][][2]{ Bifurcation parameter, $\kappa$}
\psfrag{0.95}[][][2]{\scriptsize $0.95$}
\psfrag{0.96}[][][2]{\scriptsize $0.96$}
\psfrag{0.97}[][][2]{\scriptsize $0.97$}
\psfrag{0.98}[][][2]{\scriptsize $0.98$}
\psfrag{0.99}[][][2]{\scriptsize $0.99$}
\psfrag{1.00}[][][2]{\scriptsize $1.00$}
\psfrag{1.01}[][][2]{\scriptsize $1.01$}
\psfrag{1.02}[][][2]{\scriptsize $1.02$}
\psfrag{1.03}[][][2]{\scriptsize $1.03$}
\psfrag{1.04}[][][2]{\scriptsize $1.04$}
\psfrag{1.05}[][][2]{\scriptsize $1.05$}
\psfrag{0}[][][2]{\scriptsize $0$}
\psfrag{10}[][][2]{\scriptsize $10$}
\psfrag{20}[][][2]{$20$}
\psfrag{30}[][][2]{$30$}
\psfrag{40}[][][2]{$40$}
\psfrag{4}[][][2]{\scriptsize $4$}
\psfrag{2}[][][2]{\scriptsize$2$}
\psfrag{5}[][][2]{\scriptsize $5$}
\psfrag{60}[][][2]{$60$}
\psfrag{104}{\small{$\times 10^{-4}$}}
\includegraphics[scale = 0.9,trim=0cm 0cm 0cm 1.7cm, clip=true,width=3.25in,height=2.25in]{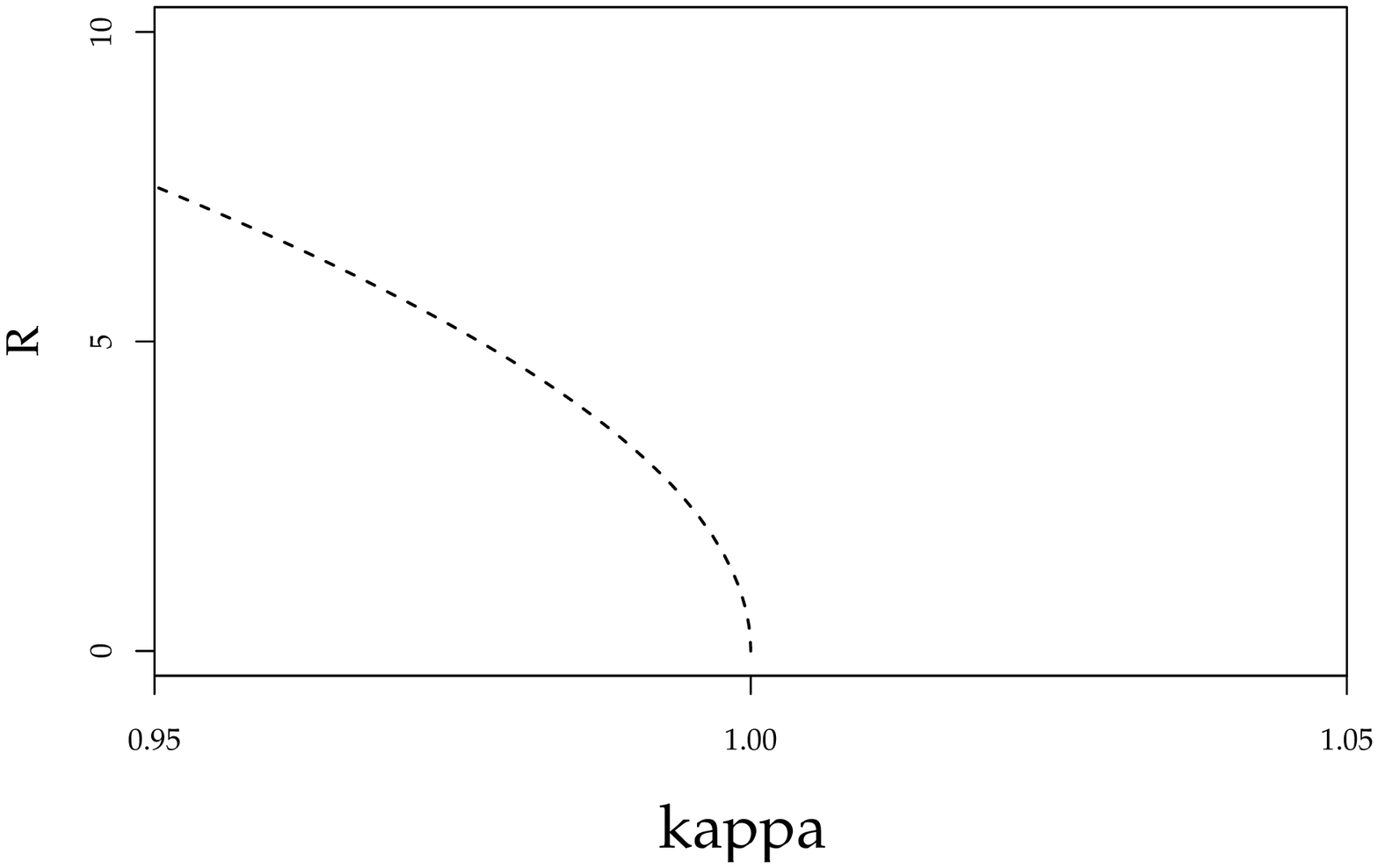}
\vspace{-2mm}
\caption{Bifurcation diagram showing the existence of a sub-critical Hopf for $a=0.95$, $b=0.022$, $C=100$, $\tau_1=10$ and $\tau_2=20$.}
%\vspace{-5mm}
\label{fig:rcp2d_withq_bfd_wrt_util_sub}
\end{figure}

\begin{figure}[hbtp!]
\psfrag{time}{\hspace{-0.1cm}  \scriptsize Time}
\psfrag{Rate}{\hspace{-0.7cm}\vspace{2cm} Rate, $R(t)$}
\psfrag{0}{\hspace{-0.04cm}\scriptsize{$0$}}
\psfrag{5}{\hspace{-0.05cm}\scriptsize{$5$}}
\psfrag{10}{\hspace{-0.1cm}\scriptsize{$10$}}
\psfrag{75}{\hspace{-0.1cm}\scriptsize{$75$}}
\psfrag{30}{\hspace{-0.1cm}\scriptsize{$30$}}
\psfrag{25}{\hspace{0cm}\scriptsize{$25$}}
\psfrag{50}{\hspace{-0.1cm}\scriptsize{$50$}}
\psfrag{100}{\hspace{-0.1cm}\scriptsize{$100$}}
\psfrag{200}{\hspace{-0.1cm}\scriptsize{$200$}}
\psfrag{0.5}{\hspace{-0.1cm}\scriptsize{$0.95$}}
\psfrag{1.0}{\hspace{-0.1cm}\scriptsize{$1.0$}}
\psfrag{1.5}{\hspace{-0.1cm}\scriptsize{$1.05$}}
\psfrag{150}{\hspace{-0.1cm}\scriptsize{$150$}}
\psfrag{300}{\hspace{-0.1cm}\scriptsize{$300$}}
\psfrag{400}{\hspace{-0.1cm}\scriptsize{$400$}}
\psfrag{-800}{\hspace{-0.1cm}\scriptsize{$-800$}}
\psfrag{-400}{\hspace{-0.1cm}\scriptsize{$-400$}}
\psfrag{-300}{\hspace{-0.1cm}\scriptsize{$-300$}}
\psfrag{-600}{\hspace{-0.1cm}\scriptsize{$-600$}}
\psfrag{125}{\hspace{-0.1cm}\scriptsize{$125$}}
\psfrag{1500}{\hspace{-0.1cm}\scriptsize{$1500$}}
\psfrag{60}{\hspace{-0.1cm}\scriptsize{$60$}}
\psfrag{120}{\hspace{-0.1cm}\scriptsize{$120$}}
\psfrag{2500}{\hspace{-0.1cm}\scriptsize{$2500$}}
\psfrag{500}{\hspace{-0.1cm}\scriptsize{$500$}}
\psfrag{1000}{\hspace{-0.1cm}\scriptsize{$1000$}}
\psfrag{2100}{\hspace{-0.1cm}\scriptsize{$2000$}}
\psfrag{3000}{\hspace{-0.1cm}\scriptsize{$3000$}}
\begin{tabular}{c}
\subfloat[ $\kappa = 0.95$]{\includegraphics[width=\subwidth]{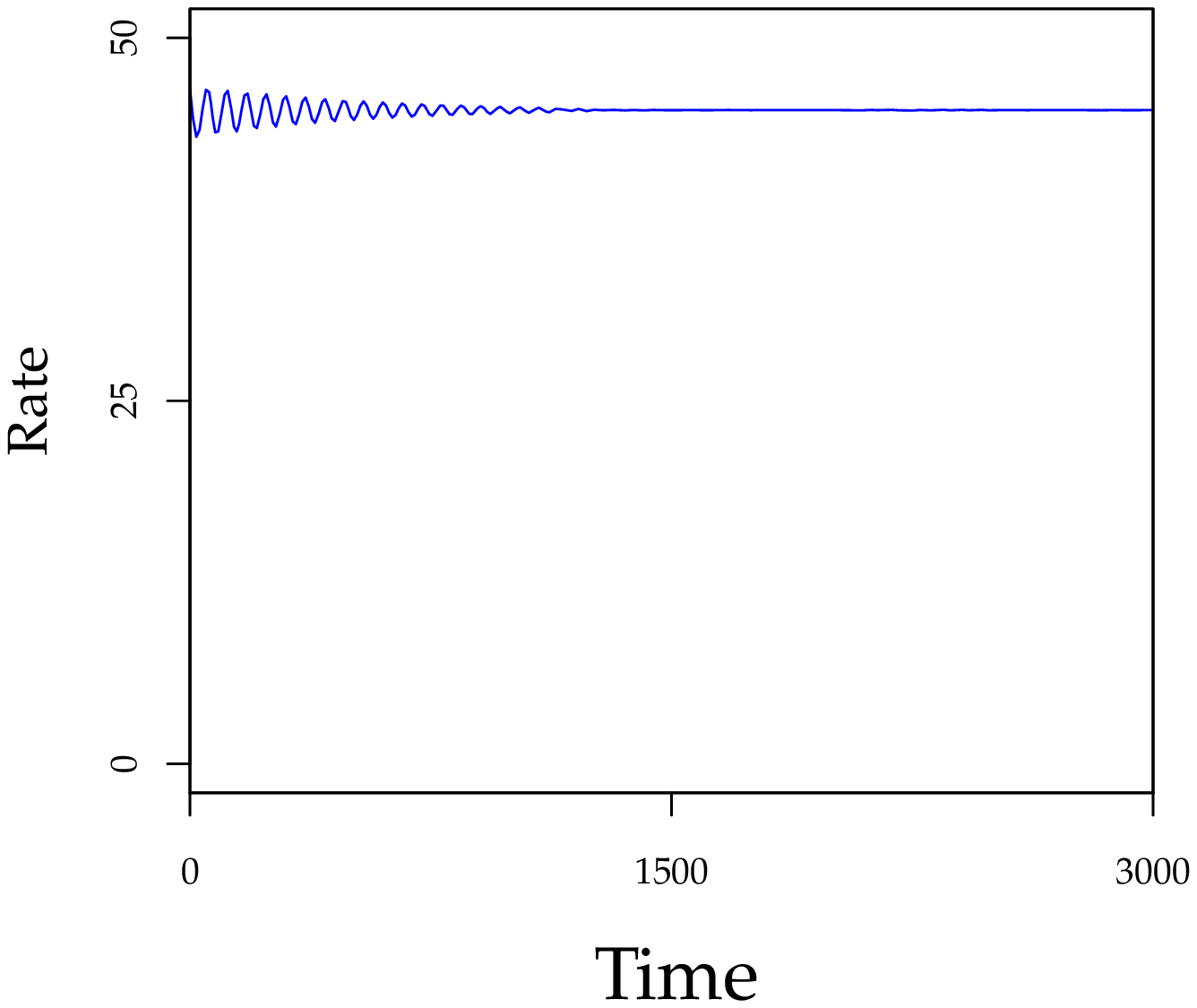}}\hspace{-5mm}
% \subfloat[ $\kappa = 0.97, R_0=2$]{\includegraphics[width=\subwdth]{tompecssub2.eps}}\hspace{-5mm}
\subfloat[ $\kappa = 1.05$]{\includegraphics[width=\subwidth]{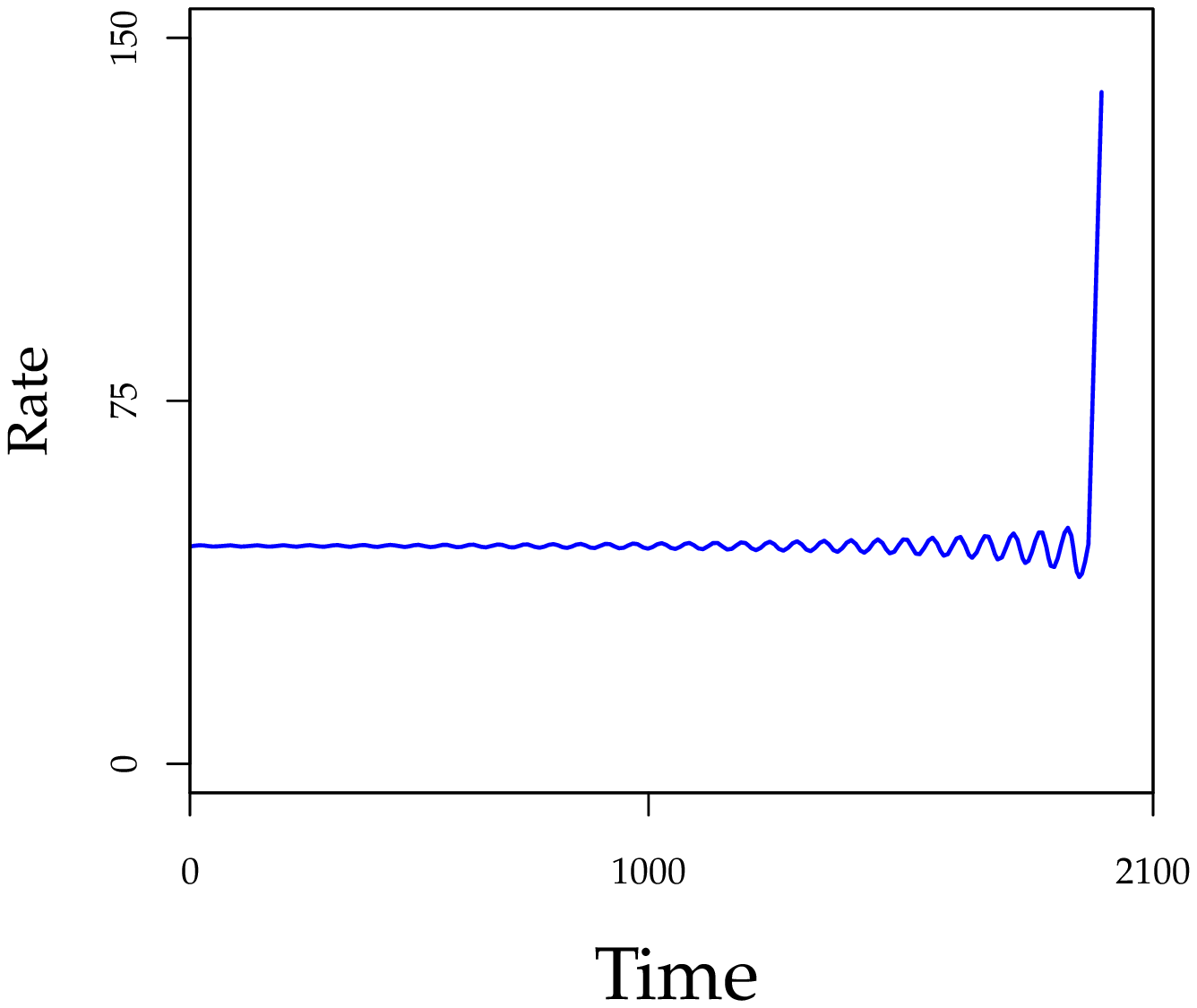}}
\end{tabular}
\caption{Numerical simulations highlighting that the system undergoes a sub-critical Hopf for the parameter values $\kappa_c= 1$, $a=0.95$, $\tau=100$, $C=10$ and $b=0.022$.} \label{fig:rcp2d_withq_ns_wrt_util_sub}
\end{figure}

$\textit{Numerical Example 1 (Super-critical)}$: Let us consider the RCP system with $C=100$, $\tau_1=10$ and $\tau_2=20$ which corresponds to $\vartheta = \pi/3$. We set $b=0.736$ which corresponds to equilibrium utilization of 55\% of link capacity i.e., $\rho^*=0.55$. For these values, using \eqref{eq:nsstability_relation}, we obtain $a=1.17$, for $\kappa_c =1$. From Figure \ref{fig:myu2Vsrho}, we can observe that $ \mu_2 > 0$ for  $\rho^*=0.55$, implying that the system undergoes a super-critical Hopf bifurcation. The bifurcation diagram drawn using the Matlab package DDE-Biftool is shown in Figure \ref{fig:rcp2d_bfd_myuvsutil_sup}. As expected, it shows that the system loses local stability via a super-critical Hopf bifurcation, as the bifurcation parameter crosses the critical threshold ($\kappa_c =1$). To validate this, numerical simulations obtained using XPPAUT are shown in Figure \ref{fig:rcp2d_withq_ns_wrt_util_super}. For $\kappa=0.95$, the system converges to the equilibrium rate, $R^*=27.5$ (see Figure \ref{fig:rcp2d_withq_ns_wrt_util_super}(a)). Whereas, for $\kappa = 1.05 > \kappa_c$ i.e. after the bifurcation, the system leads to the emergence of stable limit cycles.\\
%%%%%%%%%%%%%%%%%%%%%%%%%%%%
%\newcommand{\supwdth}{0.5\textwidth}
\\
%%%%%%%%%%%%%%%%%%%%%%%%%%%%%%%%%%%%%%%%%%%%%%%%%%%%%
\\
$\textit{Numerical Example 2 (Sub-critical)}$: Consider $a=0.95$, $b=0.022$ ($\rho^*=0.9$), $C=100$, $\tau_1=10$ and $\tau_1=20$. The system undergoes a Hopf bifurcation at $\kappa$ = 1. For these values, from Figure \ref{fig:myu2Vsrho}, we get $\mu_2 < 0$, implying that the system undergoes a sub-critical Hopf. We can also observe from Figure \ref{fig:rcp2d_withq_bfd_wrt_util_sub} that there exists no stable limit cycle in the neighborhood, as the bifurcation parameter is varied beyond the critical threshold. Hence, the Hopf bifurcation is sub-critical and the bifurcating limit cycles are unstable.
To illustrate the occurrence of a sub-critical Hopf, we present some numerical simulations in Figure \ref{fig:rcp2d_withq_ns_wrt_util_sub}. For $\kappa=0.95$, the system converges to the stable equilibrium, $R^*=45$ (see Figure \ref{fig:rcp2d_withq_ns_wrt_util_sub}(a)). Whereas, after the bifurcation i.e. for $\kappa > \kappa_c$, the previously stable fixed point now becomes unstable and also the solution would eventually jump to infinity (Figure \ref{fig:rcp2d_withq_ns_wrt_util_sub}(b)).
%%%%%%%%%%%%%%%%%%%%%%%%%%%%%%%%%%%%%%%%%%%%%%%%%%%%%%
%%%%%%%%%%%%%%%%%%%%%%%%%%%%%%%%%%%%%%%%%%%%%%%%%%%%%%%%%%%%%%%%%%%%%%%%%%%%%%%%%%%%%%%%%%%%%%%%%%%%%%%%%%%%%%%%%%%%%%%%%%%%%%%%%

In summary, the results of theoretical and numerical analysis reveal that the RCP which uses both rate mismatch and queue size feedback, can undergo a sub-critical Hopf bifurcation, which is undesirable for engineering applications. In fact, in the context of congestion control algorithms, the possibility of occurrence of a sub-critical Hopf has not been extensively studied so far.
%The Hopf bifurcation analysis reveals some key insights into the dynamical properties of RCP. 
The insights from Hopf bifurcation analysis could guide design considerations such that any loss of local stability only occurs via the emergence of small amplitude stable limit cycles. In other words, the nature of Hopf bifurcation and the stability of the bifurcating limit cycles should also be considered while designing congestion control protocols. %Therefore, this study signals the need for additional studies to understand more 

\subsection{Packet-level simulations}
We now validate some of the theoretical insights by investigating if the packet-level simulations of the underlying system exhibits the qualitative properties predicted through the Hopf bifurcation analysis of the fluid model. The packet-level simulations are done using a discrete event RCP simulator (for more details, refer to \citep{krv2009}). The simulated network has a single bottleneck link setup that considers Capacity, $C= 1$ Giga bits per sec (Gbps) and number of sources = $100$.

\begin{figure}[hbtp!]
 \psfrag{60}{\hspace{-0.2cm}\small{$50000$}}
\psfrag{0}{\small{$0$}} 
 \psfrag{30}{\hspace{0.1cm}\small{$0$}}
 \psfrag{50}{\hspace{-0.2cm}\small{$20000$}}
\psfrag{40}{\hspace{-0.2cm}\small{$10000$}}
\psfrag{15}{\hspace{-0.2cm}\small{$5000$}}
\psfrag{250}{\small{$250$}}
\psfrag{500}{\small{$500$}}
\psfrag{300}{\small{$300$}}
\psfrag{600}{\small{$600$}}
\psfrag{1000}{\small{$1000$}}
\psfrag{1500}{\small{$1500$}}
\psfrag{2000}{\small{$2000$}}
\psfrag{4000}{\small{$4000$}}
\psfrag{10000}{\small{$10000$}}
\psfrag{20000}{\small{$20000$}}
\psfrag{750}{\small{$750$}}
\psfrag{15000}{\small{$15000$}}
\centering
\begin{tabular}{cc} 
	\hspace{0.3cm}\small{Queue Size (packets)} & \hspace{0.3cm}\small{Rate (bytes/ms)} \\
	\includegraphics[height=\hite, width=0.45\textwidth]{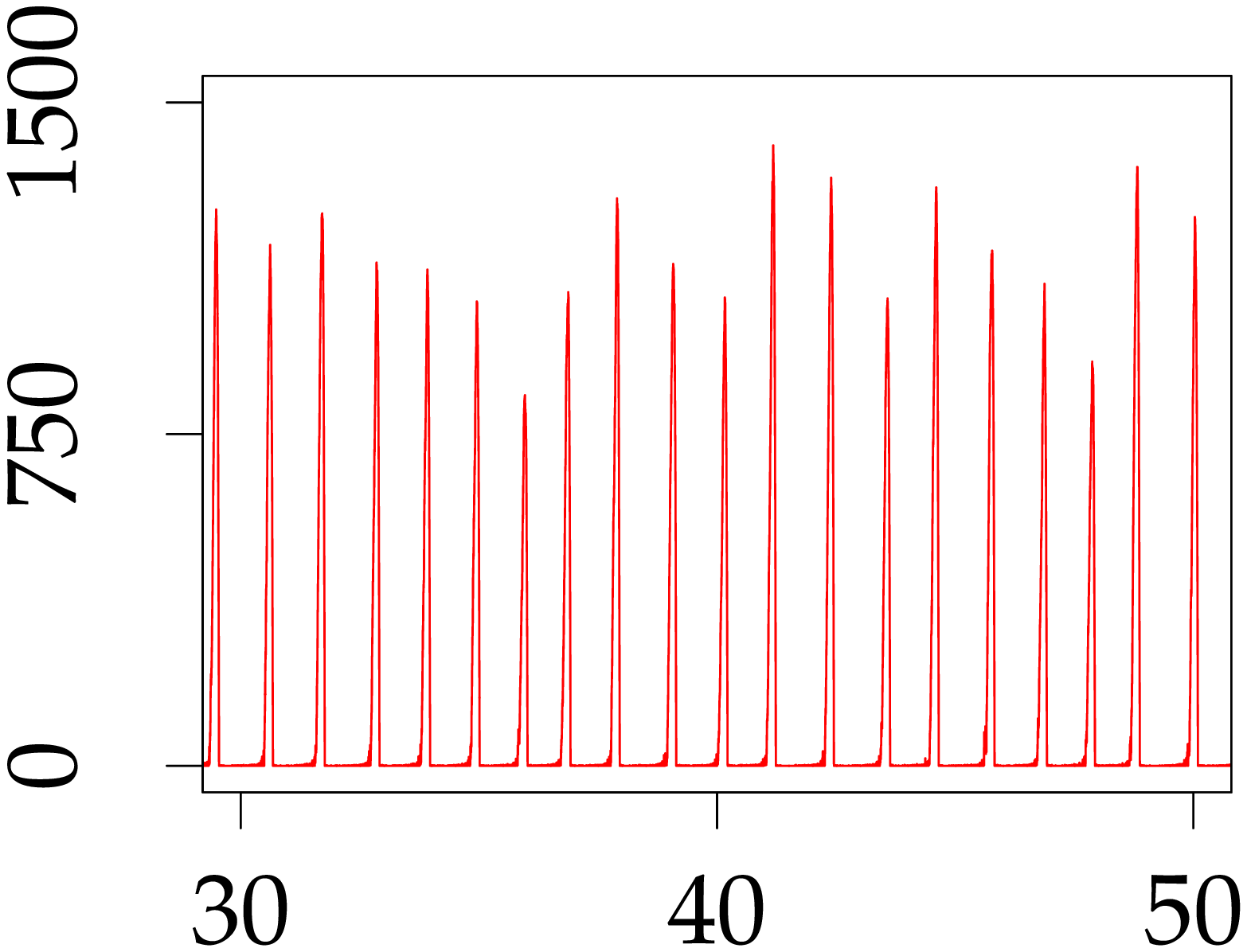} &
	\includegraphics[height=\hite, width=0.45\textwidth]{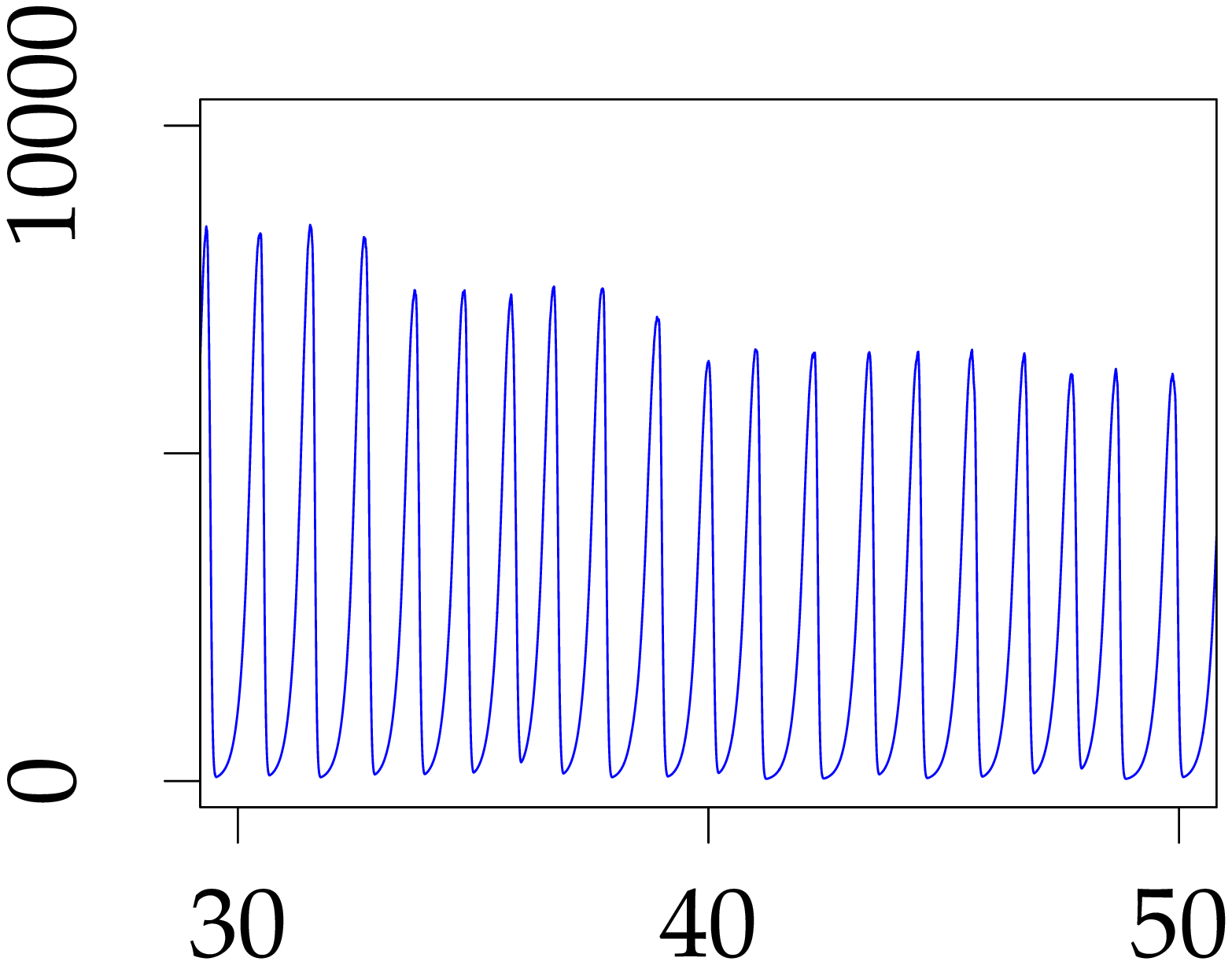} \\
 \multicolumn{2}{c}{\small{With queue feedback: $a=0.85$ and $b=0.005$}}\\
\includegraphics[height=\hite, width=0.45\textwidth]{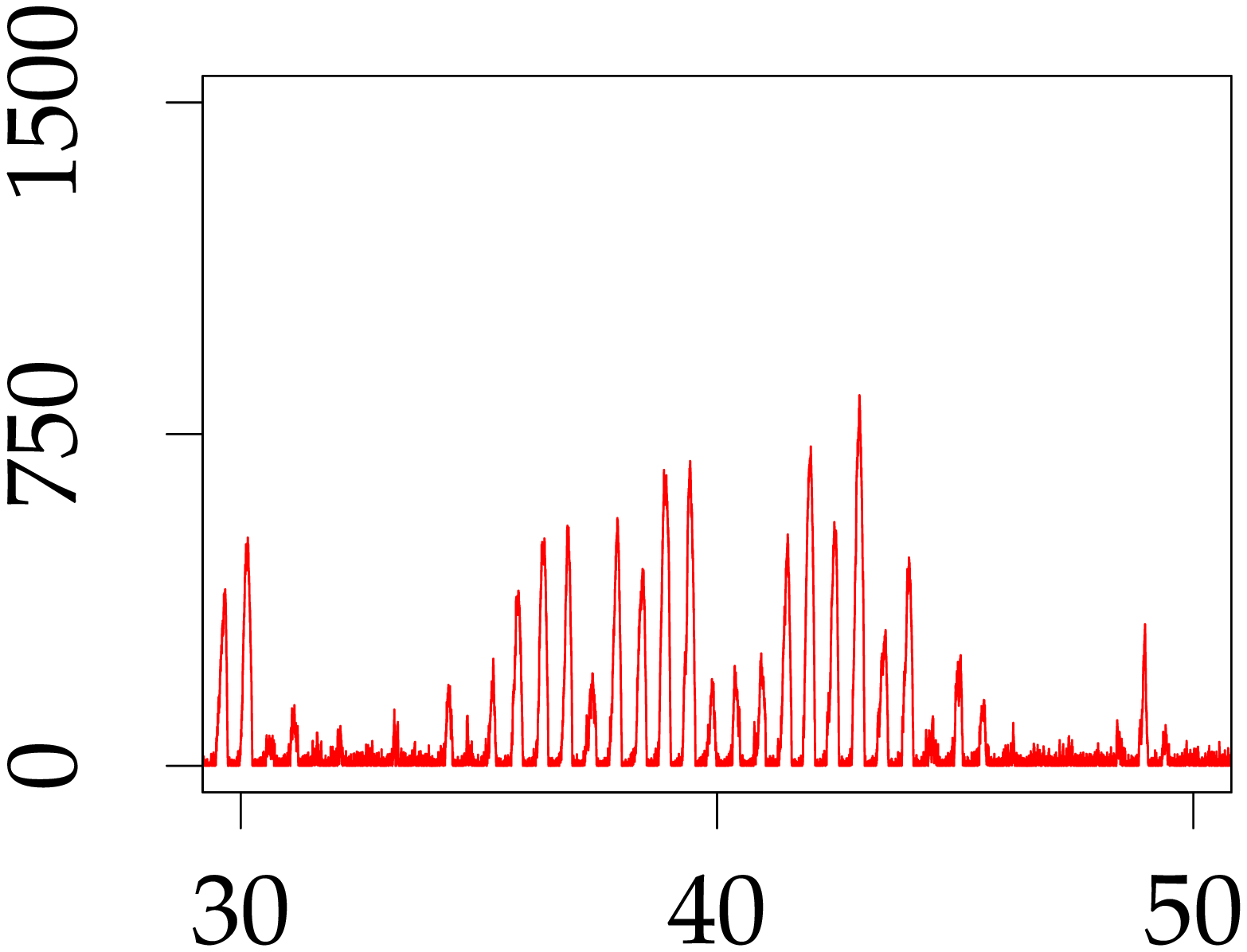} &
\includegraphics[height=\hite, width=0.45\textwidth]{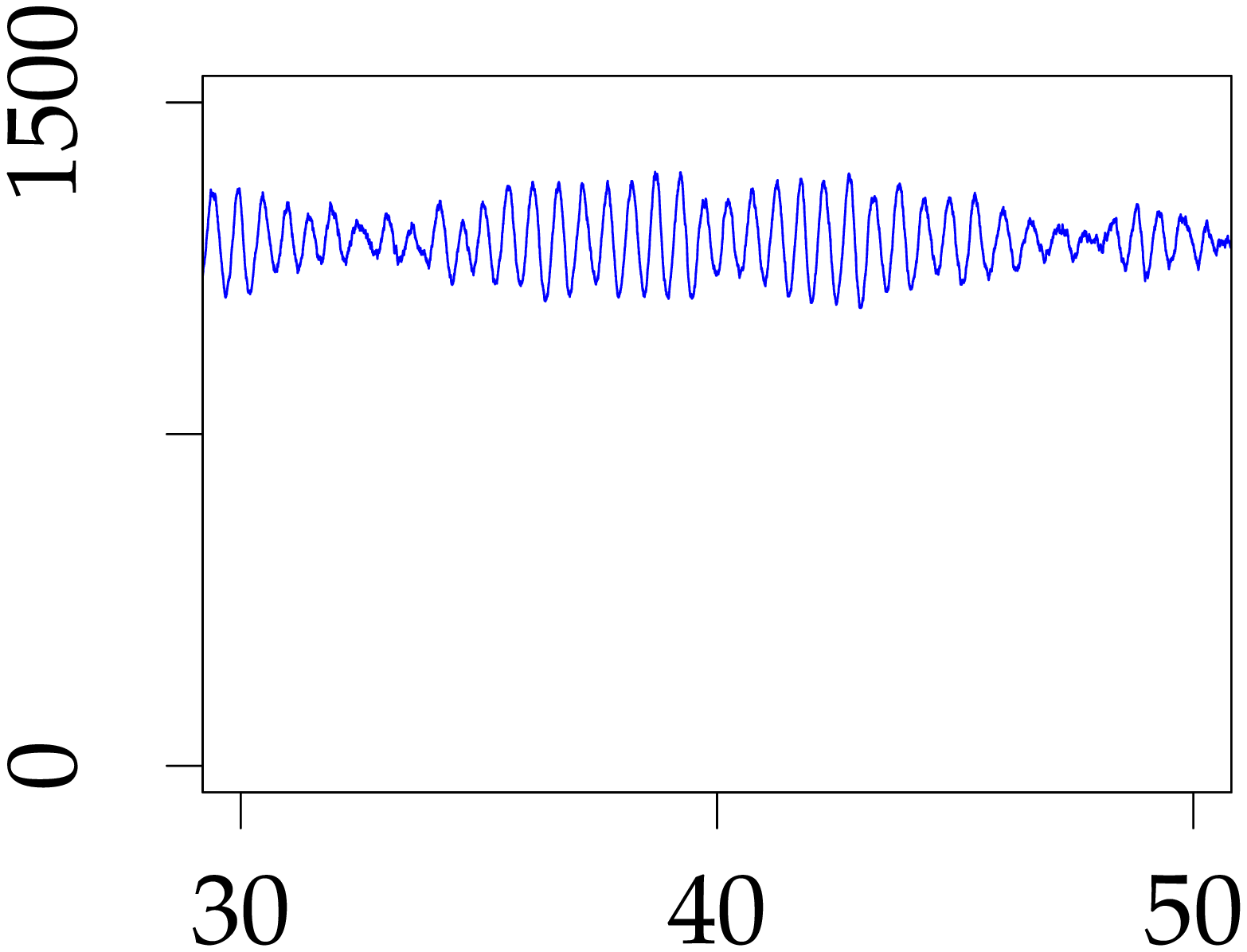} \\
 \multicolumn{2}{c}{\small{Without queue feedback: $a=1.6$, $b=0$ and $\gamma=0.95$}}\\
% 	\includegraphics[height=\hite,width=0.45\textwidth]{pl_sims_rcp_1d_hba_q_noq/100mbps/100_100_0.8000_0.0029_1.0_0/queue_size.eps} &
% 	\includegraphics[height=\hite,width=0.45\textwidth]{pl_sims_rcp_1d_hba_q_noq/100mbps/100_100_0.8000_0.0029_1.0_0/rate.eps} \\
%  \multicolumn{2}{c}{\small{With queue feedback: $C=100$ Mbps}}\\
% 	\includegraphics[height=\hite,width=0.45\textwidth]{pl_sims_rcp_1d_hba_q_noq/100mbps/100_100_1.6_0_0.95_0/queue_size.eps} &
% 	\includegraphics[height=\hite,width=0.45\textwidth]{pl_sims_rcp_1d_hba_q_noq/100mbps/100_100_1.6_0_0.95_0/rate.eps} \\
%  \multicolumn{2}{c}{\small{Without queue feedback: $C=100$ Mbps}}\\ 
	\multicolumn{2}{c}{\small{Time (ms)}}\\
\end{tabular}
\caption{\small Simulation traces highlighting that the system which includes queue feedback exhibits limit cycles with amplitude much larger than that of RCP which uses only rate mismatch feedback. The parameter values used are i) with queue feedback: $a=0.85$, $b= 0.005$ and ii) without queue feedback: $a=1.6$, $\gamma = 0.95$. We consider round-trip times of $\tau_1=100$ ms and $\tau_2=150$ ms.}
\label{fig:rcp2dhbaplsims_qVsnoq}
\end{figure}

Simulation traces in Figure \ref{fig:rcp2dhbaplsims_qVsnoq} show the evolution of queue size and flow rate for the cases with and without queue size feedback. We choose $\tau_1=100$ ms for half of the flows and $\tau_2=150$ ms for the remaining $50$ flows. Here, we set $b=0.005$ which corresponds to equilibrium utilization of 95\% of link capacity. For RCP without queue size feedback, we set $\gamma=0.95$ to achieve the same target link utilization. We can observe that the RCP which uses both rate mismatch and queue feedback exhibits large amplitude limit cycles (due to the occurrence of a sub-critical Hopf). Whereas, in the absence of queue feedback, the system undergoes a super-critical Hopf bifurcation, and leads to the emergence of small amplitude limit cycles.

\begin{figure}[hbtp!]
 \psfrag{60}{\hspace{-0.2cm}\small{$50000$}}
\psfrag{0}{\small{$0$}} 
 \psfrag{30}{\hspace{0.1cm}\small{$0$}}
 \psfrag{31}{\hspace{-0.2cm}\small{$1000$}}
\psfrag{32}{\hspace{-0.2cm}\small{$2000$}}
\psfrag{15}{\hspace{-0.2cm}\small{$5000$}}
\psfrag{250}{\small{$250$}}
\psfrag{500}{\small{$500$}}
\psfrag{300}{\small{$300$}}
\psfrag{600}{\small{$600$}}
\psfrag{1000}{\small{$1000$}}
\psfrag{1500}{\small{$1500$}}
\psfrag{2000}{\small{$2000$}}
\psfrag{4000}{\small{$4000$}}
\psfrag{12000}{\small{$12000$}}
\psfrag{60000}{\small{$60000$}}
\psfrag{7500}{\small{$7500$}}
\psfrag{15000}{\small{$15000$}}
\centering
\begin{tabular}{cc} 
	\hspace{0.3cm}\small{Queue Size (packets)} & \hspace{0.3cm}\small{Rate (bytes/ms)} \\
	\includegraphics[height=\hite, width=0.45\textwidth]{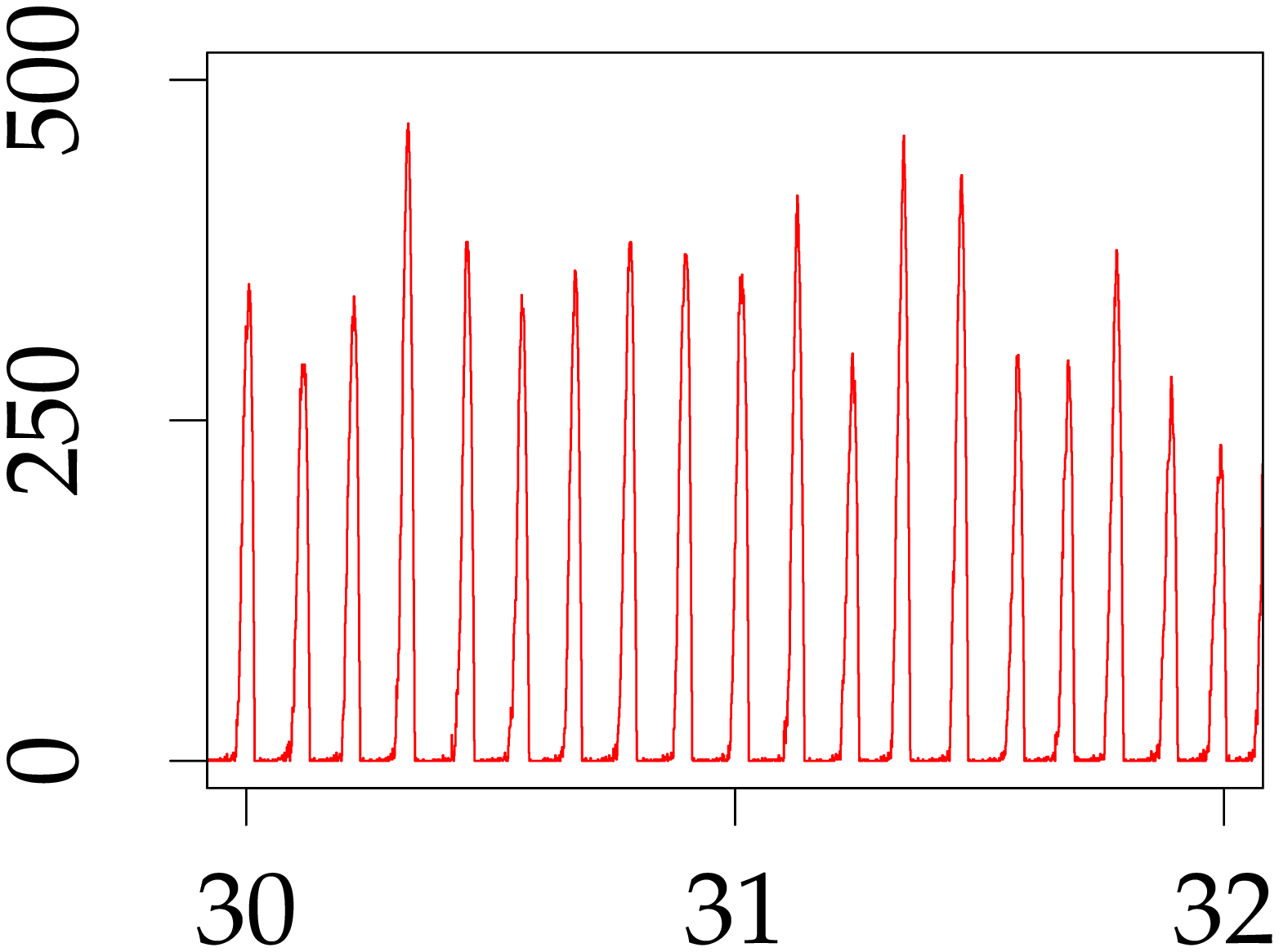} &
	\includegraphics[height=\hite, width=0.45\textwidth]{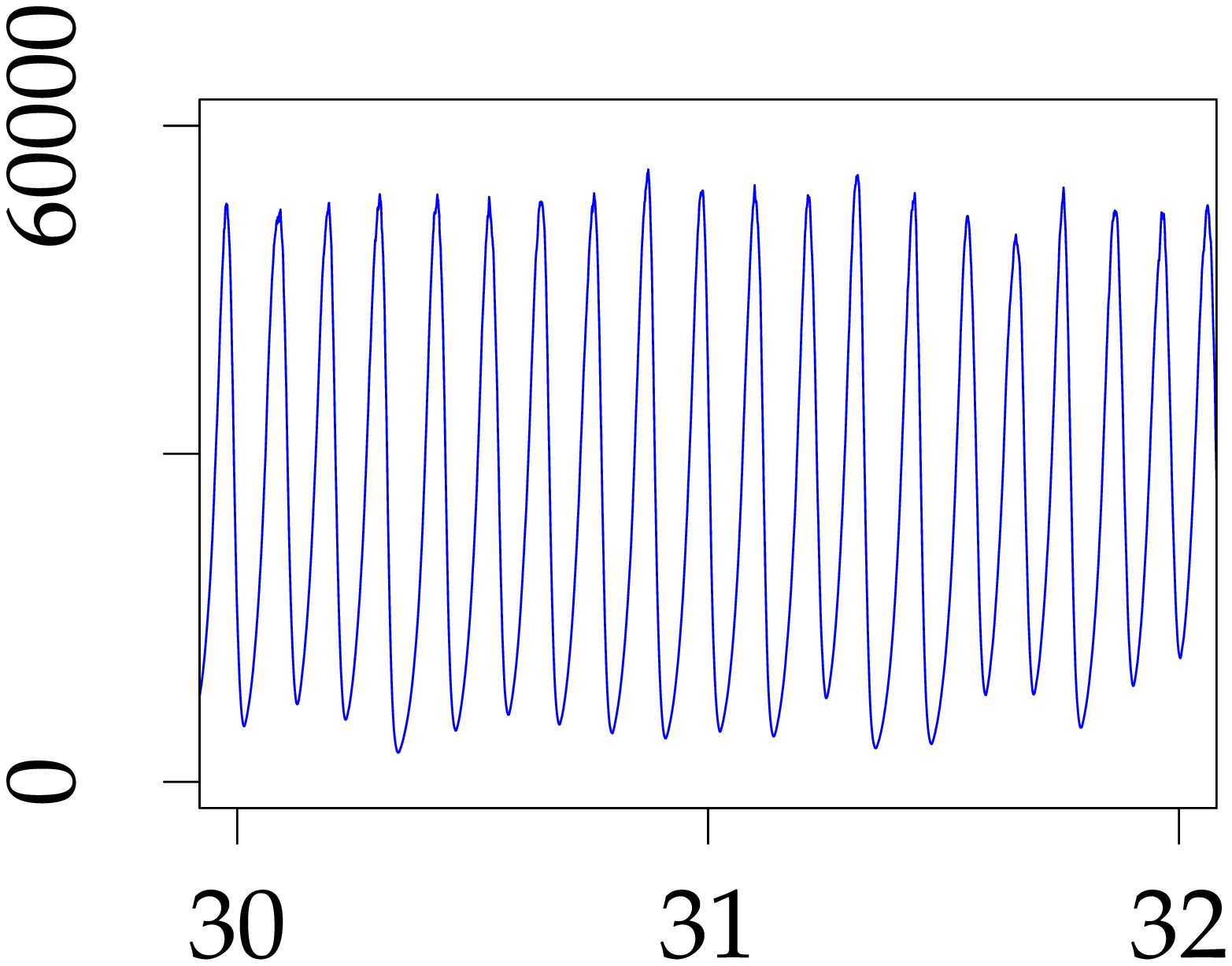} \\
 \multicolumn{2}{c}{\small{$a=1$, $\tau_1=10$ ms and $\tau_2=20$ ms}}\\
\includegraphics[height=\hite, width=0.45\textwidth]{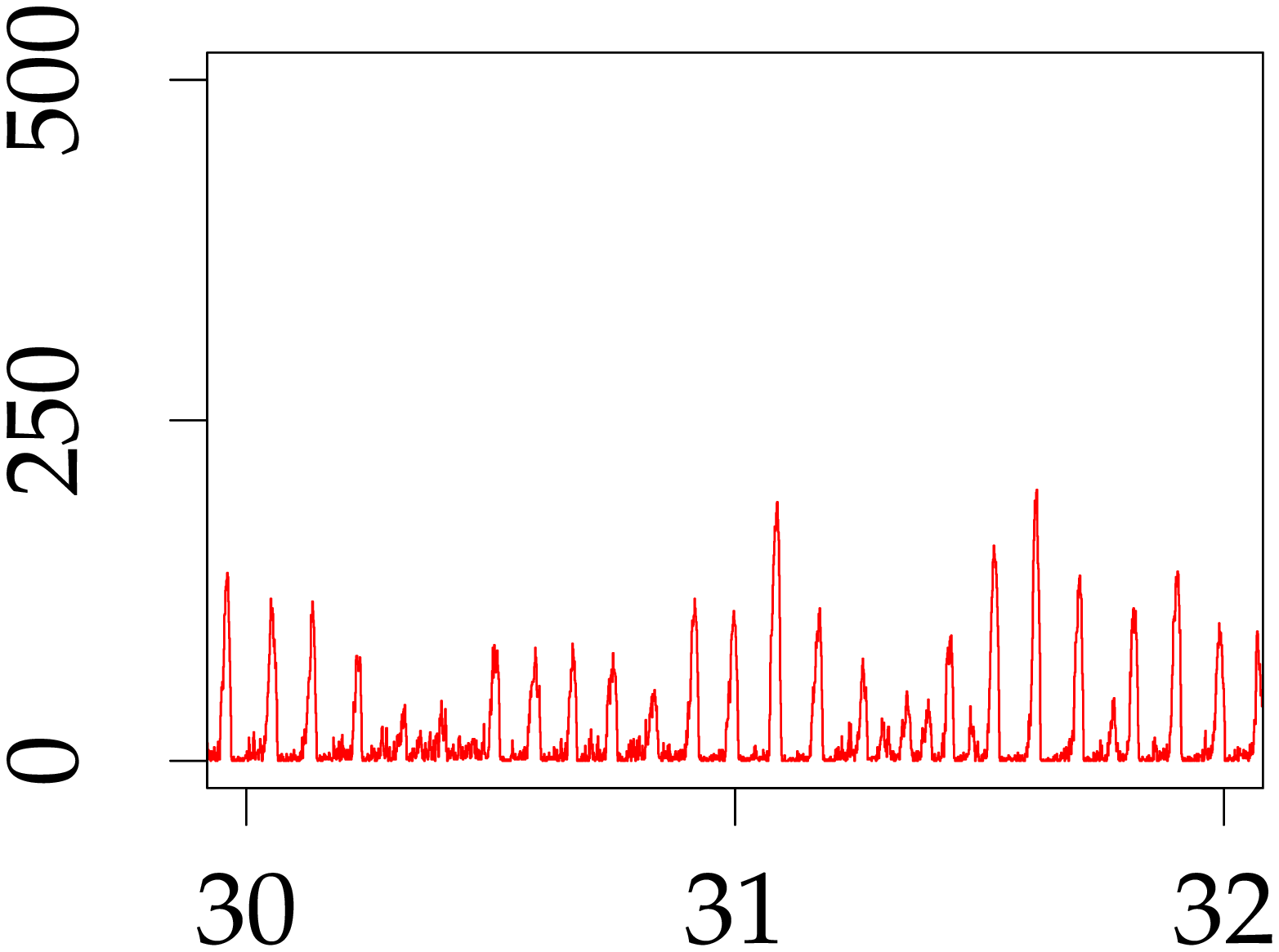} &
\includegraphics[height=\hite, width=0.45\textwidth]{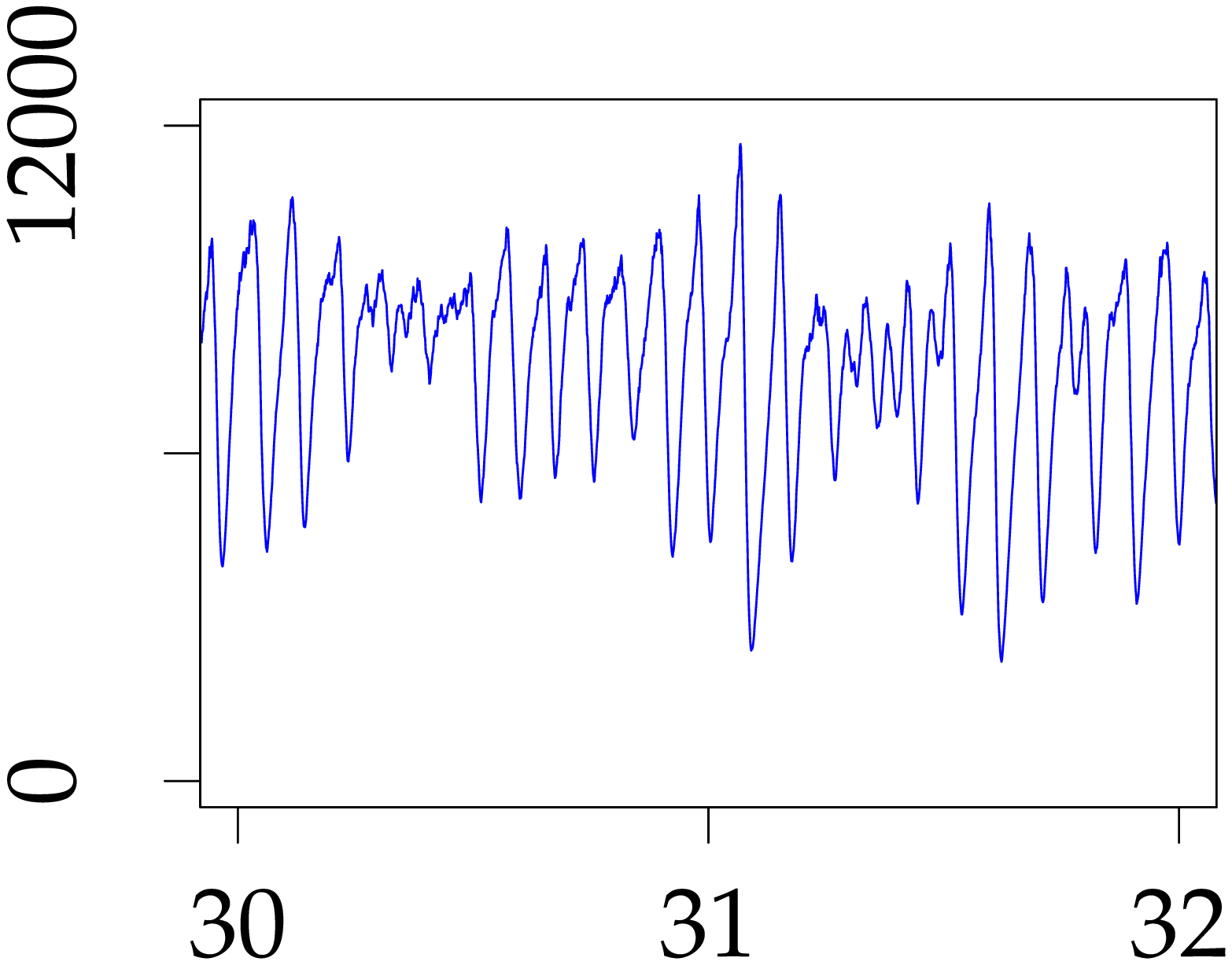} \\
 \multicolumn{2}{c}{\small{$a=1.6$, $\tau_1=10$ ms and $\tau_2=50$ ms}}\\
% 	\includegraphics[height=\hite,width=0.45\textwidth]{pl_sims_rcp_1d_hba_q_noq/100mbps/100_100_0.8000_0.0029_1.0_0/queue_size.eps} &
% 	\includegraphics[height=\hite,width=0.45\textwidth]{pl_sims_rcp_1d_hba_q_noq/100mbps/100_100_0.8000_0.0029_1.0_0/rate.eps} \\
%  \multicolumn{2}{c}{\small{With queue feedback: $C=100$ Mbps}}\\
% 	\includegraphics[height=\hite,width=0.45\textwidth]{pl_sims_rcp_1d_hba_q_noq/100mbps/100_100_1.6_0_0.95_0/queue_size.eps} &
% 	\includegraphics[height=\hite,width=0.45\textwidth]{pl_sims_rcp_1d_hba_q_noq/100mbps/100_100_1.6_0_0.95_0/rate.eps} \\
%  \multicolumn{2}{c}{\small{Without queue feedback: $C=100$ Mbps}}\\ 
	\multicolumn{2}{c}{\small{Time (ms)}}\\
\end{tabular}
\caption{\small Simulation traces highlighting that the system which includes queue feedback exhibits both super-critical and sub-critical Hopf, depending on the values of RTTs. We set $b=0.005$ which corresponds to the target utilization of $95$\% of link capacity.}
\label{fig:rcp2dhbaplsims_criticVstau}
\end{figure}

From the simulation traces shown in Figure \ref{fig:rcp2dhbaplsims_criticVstau}, we can observe that, in the presence of queue feedback, the type of Hopf bifurcation can be varied by changing the values of $\tau_1$ and $\tau_2$. 
% The results of Hopf bifurcation analysis revealed that the type of Hopf bifurcation does not depend on the values of link capacity, round-trip time and the number of flows.  We can verify this by changing the values of these parameters and observe the results in both the cases i.e., with and without queue size feedback (see Figures \ref{fig:rcp1dhbaplsims_set1} and \ref{fig:rcp1dhbaplsims_set2}).

%%%%%%%%%%%%%%%%%%%%%%%%%%%%%%%%%%%%%%%

\section{Contributions}
\label{sec:rcp2dconcl}
RCP estimates the fair rate of flows using feedback based on rate mismatch and queue size. An open design question in RCP is whether it is advantageous to include queue size feedback, given that the protocol already includes feedback based on rate mismatch. To address this question, we analyzed stability and Hopf bifurcation properties for both the design options, i.e., with and without queue size feedback. In this paper, we considered the RCP model that assumes single bottleneck link carrying flows with two different round-rip times. In the local stability analysis, we showed that in the presence of queue feedback, the system readily loses its stability via a Hopf bifurcation, as the bifurcation parameter varies. For our analysis, a dimensionless exogenous parameter was used as the bifurcation parameter. 
%Whereas, the results of rate of convergence analysis do not provide any design guidelines on whether the queue size feedback is useful or not. 
%This provides motivation for non-linear analysis to study some additional dynamical properties. 
Then, we proceeded to analyze the dynamics of both the design choices as conditions for stability are just violated. From a  bifurcation theoretic perspective we would like our algorithms to always produce stable limit cycles of small amplitude. We analyzed the type of Hopf bifurcation and the orbital stability of the bifurcating limit cycles. We highlighted that the presence of queue feedback in RCP results in a sub-critical Hopf bifurcation, for some parameter values. A sub-critical Hopf leads to either large amplitude limit cycles or unstable limit cycles, and hence its occurrence should be avoided. Whereas, in the absence of queue feedback, the Hopf bifurcation is always super-critical and leads to the emergence of stable limit cycles of small amplitude. Hence, it is advisable to go with the design choice that uses only rate mismatch feedback. We complemented the analysis with bifurcation diagrams, numerical computations, and packet-level simulations.

Naturally, the work should also extend to consider the cases with multi bottleneck link. It is also worth investigating the global stability of RCP both in the presence and absence of queue feedback.

\appendix
\section{Hopf Bifurcation Analysis}

\subsection{Existence of Hopf bifurcation}
Stability analysis highlights that the RCP loses stability at a critical value of bifurcation parameter $\kappa$; $\kappa$=$\kappa_c$. Therefore, we obtain the critical value of the bifurcation parameter, at which the system loses local stability, as
\begin{equation}
\label{eq:kappa_cval} 
\kappa_c=\dfrac{\pi}{2\myk(\tau_1+\tau_2)\cos\left(\dfrac{\omega_0(\tau_1-\tau_2)}{2}\right)}.
\end{equation}

To show that the system undergoes a Hopf bifurcation at $\kappa_c$, we need to satisfy the following transversality condition of the Hopf spectrum~\cite{hassard1981}\\
$$\mathbf{Re}\left(\dfrac{d\lambda}{d\kappa}\right)_{\kappa=\kappa_c}\neq 0.$$\\

% Differentiating (\ref{eq:chareq}) with respect to $\kappa$, and using (\ref{eq:kappa_cval}), 
% it can be shown that\\
% $$\mathbf{Re}\left(\dfrac{d\lambda}{d\kappa}\right)_{\kappa=\kappa_c} > 0.$$
% 
% 
% At $\kappa_c$, the system transits from stable to unstable regime. Now, we show that, at bifurcation point i.e. $\kappa_c$ satisfies \textit{transversality condition} of Hopf spectrum  
% $$Re\left(\dfrac{d\lambda}{d\kappa}\right)_{\kappa=\kappa_c}\neq 0.$$
% The \textit{necessary and sufficient} for the system to be stable is:
% \begin{equation} 
% \kappa\myk(\tau_1+\tau_2)\sin\left(\omega_0\tau_1\right)<\pi/2. 
% \end{equation}
% Therefore the system loses stability at $\kappa=\kappa_c$, where $$\kappa_c=\dfrac{\omega_0}{2\sin\left(\omega_0\tau_1\right)\myk}$$

Now we consider the linear autonomous delay equation of the system whose corresponding characteristic equation is given by
\begin{equation}
\lambda + \kappa\myk\left(e^{-\lambda\tau_1}+e^{-\lambda\tau_2}\right) = 0.
\end{equation}
%and satisfy the transversality condition of the Hopf spectrum.\\
Differentiating the above equation with respect to $\kappa$, we get 
$$\left. \dfrac{d\lambda}{d\kappa}\right\lvert_{\kappa=\kappa_c} = \left. \dfrac{-\myk(e^{-\lambda\tau_1}+e^{-\lambda\tau_2})}{(1-\kappa \myk (\tau_1e^{-\lambda\tau_1}+\tau_2e^{-\lambda\tau_2})\big)}\right\lvert_{\kappa=\kappa_c.}$$
Substituting the values, we obtain
$$Re\left(\dfrac{d\lambda}{d\kappa}\right)_{\kappa=\kappa_c}=\dfrac{\pi \myk\sin(\omega_0\tau_1)}{A^2+B^2} > 0,$$
where 
\begin{eqnarray}
\omega_0 &=& \pi/(\tau_1+\tau_2)\nonumber\\
A &=& 1-  \dfrac{\omega_0\cos(\omega_0\tau_1)(\tau_1-\tau_2)}{2\sin(\omega_0\tau_1)}\nonumber\\
B &=& \pi/2.\nonumber
\end{eqnarray}
Now we have shown the existence of Hopf bifurcation at the edge of the stable regime.
It is important to note that the result holds true for both the cases i.e. with, and without queue feedback , as $\myk> 0$  for both.

%The Appendix contains the requisite analysis, and we present some computational results for RCP under consideration.

%{R^2 (\tau_1 + \tau_2)}
\subsection{Type and stability of Hopf bifurcation}
We have shown that the system exhibits Hopf type bifurcation in both the cases i.e. with, and without queue feedback. However a comprehensive understanding of the stability and amplitudes of the emerging limit cycles would certainly help to address the question of whether the queue feedback is useful, or not.

Here, we outline the necessary calculations to determine the type of Hopf bifurcation and the asymptotic form of the bifurcation solutions as local instability just sets in. For now, we will only be concerned with the first Hopf bifurcation.
The framework employed to address the stability of the limit cycles is the Poincar{\'e} normal form, and the center manifold theorem. 

Consider the following nonlinear delay differential equation:
\begin{equation}
\label{eq:gen_non_eq}
\dfrac{d}{dt}x(t) = \kappa f\big(x(t),x(t-\tau_1),x(t-\tau_2)\big),
\end{equation}
where $f$ has a unique equilibrium denoted by $x^*$ and $\tau_1, \tau_2 > 0$. Define $u(t) = x(t)-x^*,$ and take a Taylor expansion for  (\ref{eq:gen_non_eq}) including the linear, quadratic and cubic terms to obtain
\vspace{-1.5mm}
\begin{equation}
\begin{aligned}
\dfrac{d}{dt}u(t)&= \kappa \big(\xi_yu(t-\tau_1)+\xi_zu(t-\tau_2)+\xi_{xy}u(t)u(t-\tau_1)\\
%&+\ \xi_{xy}u(t)u(t-\tau_1)\\
&+\ \xi_{xz}u(t)u(t-\tau_2)+\xi_{yy}u^2(t-\tau_1)\\
&+\ \xi_{yz}u(t-\tau_1)u(t-\tau_2)+\xi_{zz}u^2(t-\tau_2)\\
&+\ \xi_{xyy}u(t)u^2(t-\tau_1)+\xi_{xzz}u(t)u^2(t-\tau_2)\\
&+\ \xi_{xyz}u(t)u(t-\tau_1)u(t-\tau_2)+\xi_{yyy}u^3(t-\tau_1)\\
&+\ \xi_{yyz}u^2(t-\tau_1)u(t-\tau_2)+\xi_{zzz}u^3(t-\tau_2)\\
&+\ \xi_{yzz}u(t-\tau_1)u^2(t-\tau_2)+ \mathcal{O}(u^4)\big)\\
\end{aligned}
\label{eq:linear_noneq}
\end{equation}
where, letting $f^*$ denote evaluation of $f$ at $(x^*,y^*,z^*)$
\begin{alignat*}
\xi\xi_i&=f^*_i,&\xi_{ii}&=\dfrac{1}{2}f^*_{ii},&\xi_{iii}&=\dfrac{1}{6}f^*_{iii} \quad \forall \  i \in \{x,y,z\}\\
\xi_{xy}&=f^*_{xy},& \xi_{xz}&=f^*_{xz},& \xi_{yz}&=f^*_{yz},\ \xi_{xxy}=\dfrac{1}{2}f^*_{xxy}  \\
\xi_{xxz}&=\dfrac{1}{2}f^*_{xxz}, & \  \xi_{xyy}&=\dfrac{1}{2}f^*_{xyy},&\  \xi_{xzz}&=\dfrac{1}{2}f^*_{xzz}\\
\xi_{xyz}&=f^*_{xyz}, &\xi_{yyz}&=\dfrac{1}{2}f^*_{yyz},& \xi_{yzz}&=\dfrac{1}{2}f^*_{yzz}.
\end{alignat*}

%of (\ref{eq:linear_noneq}
The calculations that follow will enable us to address questions about the form of the bifurcating solutions, as the system transits
from stability to instability via a Hopf bifurcation. For this we have to take higher order terms, i.e., the quadratic and cubic of (\ref{eq:linear_noneq}) into consideration. Following the work of~\cite{gaurav2005}, we now perform the requisite calculations.

Consider the following autonomous delay-differential system
\begin{equation}
\label{eq:auto_noneq}
\dfrac{d}{dt}u(t) = \mathcal{L}_\mu u_t + \mathcal{F}(u_t,\mu),
\end{equation}
where $t>0,\ \mu \in \mathbb{R},\ \tau =\text{max}(\tau_1,\tau_2) >0,$
\begin{equation*}
  u_t(\theta) = u(t+\theta),\  u:[-\tau,0]\rightarrow\mathbb{R},\  \theta\in[-\tau,0].
\end{equation*}
$\mathcal{L}_\mu$ is a one-parameter family of continuous linear operators defined as $\mathcal{L}_\mu: C[-\tau,0]\rightarrow\mathbb{R}$. The operator $\mathcal{F}(u_t,\mu):C[-\tau,0]\rightarrow\mathbb{R}$  contains the nonlinear terms. Further, assume that $\mathcal{F}(u_t,\mu)$ is analytic and that $\mathcal{F}$ and $\mathcal{L}_\mu$ depend analytically on the bifurcation parameter. Note that (\ref{eq:linear_noneq}) is a
type of the form (\ref{eq:auto_noneq}). The objective now is to rewrite (\ref{eq:auto_noneq}) as follow
\begin{equation}
\label{eq:auto_noneq_matrix}
 \dfrac{d}{dt}u_t  = \mathcal{A}(\mu)u_t+\mathcal{R}u_t
\end{equation}
which has $u_t$ rather than both $u$ and $u_t$. By the Riesz representation theorem, there exists a matrix-valued function $\eta(.,\mu):[-\tau,0]\rightarrow\mathbb{R}^{n^2}$, with variation of each component of $\eta$ is bounded and for all $\phi \in C[-\tau,0]$

\begin{equation*}
  \mathcal{L}_\mu\phi = \int_{-\tau}^0d\eta(\theta,\mu)\phi(\theta),
\end{equation*}
% In particular
% \begin{equation}
% \label{eq:auto_noneq1}
%   \mathcal{L}_\mu u_t = \int_{-\tau}^0d\eta(\theta,\mu)u(t+\theta).
% \end{equation}
where   $\mathrm{d}\eta(\theta,\mu) =\kappa\big(\xi_y\delta(\theta+\tau_1)+\xi_z\delta(\theta+\tau_2)\big)\mathrm{d}\theta$
% \begin{equation*}
%   d\eta(\theta,\mu) =\kappa\big(\xi_y\delta(\theta+\tau_1)+\xi_z\delta(\theta+\tau_2)\big)d\theta,
% \end{equation*}
and $\delta(\theta)$ is the Dirac delta function.

Now we define
\begin{equation}
\label{eq:Atheta}
  \mathcal{A}(\mu)\phi(\theta) =
  \begin{cases}
  \dfrac{d\phi(\theta)}{d\theta}, & \theta \in [-\tau,0),\\
  \int_{-\tau}^0d\eta(s,\mu)\phi(s), & \theta=0,
  \end{cases}
\end{equation}
and
\begin{equation*}
  \mathcal{R}\phi(\theta) =
    \begin{cases}
  0, & \theta \in [-\tau,0),\\
  \mathcal{F}(\phi,\mu),& \theta=0.
  \end{cases}
\end{equation*}
Now the system (\ref{eq:auto_noneq}) becomes equivalent to (\ref{eq:auto_noneq_matrix}) as required.

% The bifurcating periodic solutions $u(t,\mu(\epsilon))$ of (\ref{eq:auto_noneq}) (where $\epsilon \geq 0$
% is a small parameter) have amplitude $\mathcal{O}(\epsilon)$, period $\mathcal{P}(\epsilon)$
% and nonzero Floquet exponent $\beta(\epsilon)$, where $\mu,\mathcal{P}$ and $\beta$ have the
% following (convergent) expansions:
% \begin{equation*}
% \begin{aligned}
%   &\mu = \mu_2\epsilon^2+ \mu_4\epsilon^4+\cdots\\
%   &\mathcal{P} = \dfrac{2\pi}{\omega_0}(1+\mathcal{T}_2\epsilon^2+\mathcal{T}_4\epsilon^4+\cdots) \\
%   &\beta = \beta_2\epsilon^2+ \beta_4\epsilon^4+\cdots.\\
% \end{aligned}
% \end{equation*}
% The sign of $\mu_2$ determines the \emph{direction of bifurcation}: If $\mu_2>0$, the bifurcation is \emph{supercritical} and $\mu_2<0$ implies a \emph{subcritical} bifurcation. The sign of $\beta_2$ determines the stability of $u(t,\mu(\epsilon))$: \emph{asymptotic orbital stability} if $\beta_2<0$ and instability if $\beta_2>0$. These coefficients will now be determined. 
%We only need to compute the expressions at $\mu=0$, hence, we set $\mu=0$ in the following. 
Let $q(\theta)$ be the eigenfunction for $\mathcal{A}(0)$ corresponding to $\lambda(0)$, namely
\begin{equation*}
\mathcal{A}(0)q(\theta) = i\omega_0q(\theta).
\end{equation*}
Now we define an adjoint operator $\mathcal{A}^*(0)$ as
\begin{equation*}
  \mathcal{A}^*(0)\alpha(s) =
    \begin{cases}
  -\dfrac{d\alpha(s)}{ds}, & s \in (0,\tau],\\
  \int_{-\tau}^0d\eta^T(t,0)\alpha(-t),& s=0.
  \end{cases}
\end{equation*}
%where $\eta^T$ denotes the transpose of $\eta.$
\newline Note that, the domains of $\mathcal{A}$ and $\mathcal{A}^*$ are $C^1[-\tau,0]$ and $C^1[0,\tau]$ respectively. As
\begin{equation*}
  \mathcal{A}q(\theta) = \lambda(0)q(\theta)
\end{equation*}
$\bar{\lambda}(0)$ is an eigenvalue for $\mathcal{A}^*$, and
\begin{equation*}
  A^*q^* = -i\omega_0q^*
\end{equation*}
for some nonzero vector $q^*$. For $\phi \in C[-\tau,0]$ and $\psi \in C[0,\tau]$, define a bilinear inner product
\begin{equation}\label{inner_pro}
  \varsigma \langle \psi,\phi\rangle = \bar{\psi}(0).\phi(0)-\int_{\theta=-\tau}^0\int_{\varsigma=0}^\theta\bar\psi^T(\varsigma-\theta)d\eta(\theta)\phi(\varsigma)d\varsigma.
\end{equation}
Then, $\langle\psi,A \phi\rangle = \langle A^*\psi,\phi\rangle $ for $\phi \in$ Dom$(\mathcal{A}),\psi \in$ Dom$(\mathcal{A}^*)$. Let $q(\theta) = e^{i\omega_0\theta}$ and $q^*(s) = De^{i\omega_0 s}$ be the eigenvectors for $ \mathcal{A}$ and $\mathcal{A}^*$ corresponding to the eigenvalues $+i\omega_0$ and $-i\omega_0$.

%we get $\langle q^*,q\rangle = 1$ and $\langle q^*,\bar{q}\rangle = 0$.
Value of $D$ can be evaluated using (\ref{inner_pro}) and the relation $\langle q^*,q\rangle = 1$ as
\[
\begin{aligned}
 \langle q^*,q\rangle =&~  \bar{D}-\bar{D}\kappa\int_{\theta = -\tau}^0\theta e^{i\omega_0\theta}\big(\xi_y\delta(\theta+\tau_1) \\
  &~~+   \xi_z\delta(\theta+\tau_2)\big)\mathrm{d}\theta \\
 \Rightarrow 1 =&~  \bar{D}+\bar{D}\kappa\left(\tau_1\xi_ye^{-i\omega_0\tau_1}+\tau_2\xi_ze^{-i\omega_0\tau_2}\right) \\
 \Rightarrow D = &~ \frac{1}{1+\kappa\tau_1\xi_ye^{i\omega_0\tau_1}+\kappa\tau_2\xi_ze^{i\omega_0\tau_2}}. 
\end{aligned}
\]
Again, using (\ref{inner_pro}) we show that $\langle q^*,\bar{q}\rangle=0$ as
\[
\begin{aligned}\normalsize
  \langle q^*,\bar{q}\rangle = &~ \bar{D}+\frac{\bar{D}\kappa}{2i\omega_0}\int_{\theta = -\tau}^0(e^{-i\omega_0\theta}-e^{i\omega_0\theta})\nonumber\\
  &~~ \times \big(\xi_y\delta(\theta+\tau_1)+\xi_z\delta(\theta+\tau_2)\big)\mathrm{d}\theta,\nonumber\\
  =&~ \bar{D}+\frac{\bar{D}\kappa}{2i\omega_0}\big(\xi_y(e^{i\omega_0\tau_1}-e^{-i\omega_0\tau_1})\nonumber\\
  &~ + \ \xi_z(e^{i\omega_0\tau_2}-e^{-i\omega_0\tau_2})\big),\nonumber\\
  = &~ 0.\nonumber
\end{aligned}
\] 
Now we define
\begin{equation}
\begin{aligned}
z(t) &= \langle q^*.u_t\rangle  \text{and}  \\
w(t,\theta)&= u_t(\theta)-2Re\{z(t)q(\theta)\}.
\end{aligned}
\end{equation}
Then,  on the centre manifold $C_0$,\\
 $w(t,\theta)=w\big(z(t), \bar{z}(t),\theta\big)$ where
\begin{equation}\label{w_equ}
  w(z,\bar{z},\theta)=w_{20}(\theta)\frac{z^2}{2}+w_{11}(\theta)z\bar{z}+w_{02}(\theta)\frac{\bar{z}^2}{2}+\cdots.
\end{equation}
In effect, $z$ and $\bar{z}$ are local coordinates for manifold in $C$ in the directions
of $q^*$ and $\bar{q}^*$, respectively. The existence of the center
manifold $C_0$ enables us to reduce (\ref{eq:auto_noneq_matrix}) to an ordinary differential
equation for a single complex variable on $C_0$. At $\mu=0$, we have
\begin{eqnarray}
z'(t) &\hspace{-2mm}=&\hspace{-2mm} \langle q^*,\mathcal{A}y_t+\mathcal{R}u_t\rangle\nonumber\\
&\hspace{-2mm}=&\hspace{-2mm} i\omega_0z(t)+\bar{q}^*(0).\mathcal{F}\big(w(z,\bar{z},\theta)+2Re\{z(t)q(\theta)\}\big)\nonumber\\
&\hspace{-2mm}=&\hspace{-2mm} i\omega_0z(t)+\bar{q}^*(0).\mathcal{F}_0(z,\bar{z})\label{z_dot_prod}
\end{eqnarray}
which can be written as
\begin{equation}\label{abb_z}
  z'(t) = i\omega_0z(t)+g(z,\bar{z}).
\end{equation}
% The next objective is to expand $g$ in powers of $z$  and $\bar{z}$ .
% However, we also have to determine the coefficients $w_{ij}(\theta)$ in (\ref{w_equ}).
% Once the $w_{ij}$ have been determined, the differential equation (\ref{z_dot_prod}) for $z$ would be explicit [as abbreviated in (\ref{abb_z})] where 
Now expanding the function $g(z,\bar{z})$ in powers of $z$ and $\bar{z}$ we get
\begin{eqnarray}
  g(z,\bar{z}) &=& \bar{q}^*(0).\mathcal{F}_0(z,\bar{z})\nonumber\\
  &=& g_{20}\frac{z^2}{2}+g_{11}z\bar{z}+g_{02}\frac{\bar{z}^2}{2}+g_{21}\frac{z^2\bar{z}}{2}+\cdots.\nonumber
\end{eqnarray}
Following \cite{hassard1981}, we write
\begin{equation}\label{wdash}
  w' = u_t'-z'q-\bar{z}'\bar{q}.
\end{equation}
From (\ref{eq:auto_noneq_matrix}) and (\ref{abb_z}) we get
\begin{equation*}
  w' =
    \begin{cases}
  Aw-2Re\{\bar{q}^*(0).\mathcal{F}_0q(\theta)\}, & \theta \in [-\tau_2,0)\\
  Aw-2Re\{\bar{q}^*(0).\mathcal{F}_0q(0)\}+\mathcal{F}_0,& \theta=0
  \end{cases}
\end{equation*}
which can be written as
\begin{equation}\label{eq:matrix_w}
  w' = Aw+H(z,\bar{z},\theta),
\end{equation}
using (\ref{abb_z}), where
\begin{equation}\label{H_equ}
  H(z,\bar{z},\theta)=H_{20}(\theta)\frac{z^2}{2}+H_{11}(\theta)z\bar{z}+H_{02}(\theta)\frac{\bar{z}^2}{2}+\cdots.
\end{equation}
Now, on the centre manifold $C_0$, near the origin
\begin{equation*}
  w' = w_zz'+w_{\bar{z}}\bar{z}'.
 \end{equation*}
Use (\ref{w_equ}) and (\ref{abb_z}) to replace $w_z,z'$ and equating this with (\ref{eq:matrix_w}), we get
\begin{eqnarray}
  (2i\omega_0-A)w_{20}(\theta) &=& H_{20}(\theta)\label{eq1} \\
  -Aw_{11}(\theta) &=& H_{11}(\theta)\label{eq2}\\
  (2i\omega_0-A)w_{02}(\theta) &=& H_{02}(\theta)\label{eq2.1}.
\end{eqnarray}
From (\ref{wdash}), we get
\begin{eqnarray}
  u_t(\theta) &=& w(z,\bar{z},\theta)+zq(\theta)+\bar{z}\bar{q}(\theta) \nonumber\\
   & =& w_{20}(\theta)\frac{z^2}{2}+w_{11}z\bar{z}+w_{02}(\theta)\frac{\bar{z}^2}{2}\nonumber\\
  & &+ze^{i\omega_0\theta}+\bar{z}e^{-i\omega_0\theta}+\cdots\nonumber
\end{eqnarray}
from which $u_t(0)$, $u_t(-\tau_1)$ and $u_t(-\tau_2)$ can be determined. 
As  We only require the coefficients of $z^2,z\bar{z},\bar{z}^2$ and $z^2\bar{z}$ , we have
\begin{eqnarray}
u_t(-\tau_1)u_t(-\tau_2) &\hspace{-1.5mm}=&\hspace{-1.5mm} \big(w(z,\bar{z},\tau_1)+ze^{-i\omega_0\tau_1}+\bar{z}e^{i\omega_0\tau_1}\big)\nonumber\\
&\hspace{-1.5mm}&\hspace{-1.5mm} \times \big(w(z,\bar{z},\tau_2)+ze^{-i\omega_0\tau_2}+\bar{z}e^{i\omega_0\tau_2}\big)\nonumber\\
&\hspace{-1.5mm}=&\hspace{-1.5mm} z^2e^{-i\omega_0(\tau_1+\tau_2)}+\bar{z}^2e^{i\omega_0(\tau_1+\tau_2)}\nonumber\\
&\hspace{-1.5mm}&\hspace{-1.5mm} +\ z\bar{z}(e^{-i\omega_0(\tau_1-\tau_2)}+e^{i\omega_0(\tau_1-\tau_2)})\nonumber\\   
&\hspace{-1.5mm}&\hspace{-1.5mm}+\ z^2\bar{z}\left(e^{-i\omega_0\tau_1}w_{11}(-\tau_2)\right.\nonumber\\
&\hspace{-1.5mm}&\hspace{-1.5mm}\left.+\ e^{-i\omega_0\tau_2}w_{11}(-\tau_1)\right.\nonumber\\
&\hspace{-1.5mm}&\hspace{-1.5mm}\left.+\ e^{i\omega_0\tau_1}w_{20}(-\tau_2)/2\right)\nonumber\\
&\hspace{-1.5mm}&\hspace{-1.5mm}\left.+\ e^{i\omega_0\tau_2}w_{20}(-\tau_1)/2\right)+\cdots.\nonumber\\
%************************
%u_t^3(0)&\hspace{-2mm}=&\hspace{-2mm}\big(w(z,\bar{z},0)+z+\bar{z}\big)^3\nonumber\\
%&\hspace{-2mm}=&\hspace{-2mm} 3z^2\bar{z}+\cdots.\nonumber\\
%************************
%u_t^2(0)u(-\tau_j) &\hspace{-2mm}=&\hspace{-2mm}  (w(z,\bar{z},0)+z+\bar{z})^2\nonumber\\
%&\hspace{-2mm}&\hspace{-2mm}\times (w(z,\bar{z},\tau_j)+ze^{-i\omega_0\tau_j}+\bar{z}e^{i\omega_0\tau_j})\nonumber\\
%&\hspace{-2mm}=&\hspace{-2mm} z^2\bar{z}(2e^{-i\omega_0\tau_j}+e^{i\omega_0\tau_j})+\cdots;\nonumber\\&&\hspace{-2mm} \ j\in \{1,2\}.\nonumber\\
%************************
%u_t(0)u^2(-\tau_j) &\hspace{-2mm}=&\hspace{-2mm}  \big(w(z,\bar{z},0)+z+\bar{z}\big)\nonumber\\
%&\hspace{-2mm}&\hspace{-2mm}\times \big(w(z,\bar{z},\tau_j)+ze^{-i\omega_0\tau_j}+\bar{z}e^{i\omega_0\tau_j}\big)^2\nonumber\\
%&\hspace{-2mm}=&\hspace{-2mm} z^2\bar{z}(e^{-2i\omega_0\tau_j}+2)+\cdots, \ j\in \{1,2\}.\nonumber\\
%***********************
u_t(0)u_t(-\tau_1) &\hspace{-2mm}=&\hspace{-2mm} \big(w(z,\bar{z},0)+z+\bar{z}\big)\nonumber\\
\times u_t(-\tau_2)\quad&\hspace{-2mm}&\hspace{-2mm}\times \big(w(z,\bar{z},\tau_1)+ze^{-i\omega_0\tau_1}+\bar{z}e^{i\omega_0\tau_1}\big)\nonumber\\
&\hspace{-2mm}&\hspace{-2mm}\times(w(z,\bar{z},\tau_2)+ze^{-i\omega_0\tau_2}+\bar{z}e^{i\omega_0\tau_2})\nonumber\\
&\hspace{-2mm}=&\hspace{-2mm}z^2\bar{z}(e^{i\omega_0(\tau_1-\tau_2)}+e^{-i\omega_0(\tau_1-\tau_2)}\nonumber\\
&\hspace{-2mm}&\hspace{-2mm}+ e^{-i\omega_0(\tau_1+\tau_2)})+\cdots.\nonumber\\
%**********************
u_t^3(-\tau_1) &\hspace{-2mm}=&\hspace{-2mm}  \big(w(z,\bar{z},\tau_j)+ze^{-i\omega_0\tau_1}+\bar{z}e^{i\omega_0\tau_1}\big)^3\nonumber\\
&\hspace{-2mm}=&\hspace{-2mm} 3z^2\bar{z}e^{-i\omega_0\tau_1}+\cdots;\ j\in \{1,2\}\nonumber.\\
%*********************
u_t^2(-\tau_1)u_t(-\tau_2) &\hspace{-2mm}=&\hspace{-2mm} \big(w(z,\bar{z},\tau_1)+ze^{-i\omega_0\tau_1}+\bar{z}e^{i\omega_0\tau_1}\big)^2\nonumber\\
&\hspace{-2mm}&\hspace{-2mm}\times \big(w(z,\bar{z},\tau_2)+ze^{-i\omega_0\tau_2}+\bar{z}e^{i\omega_0\tau_2}\big)\nonumber\\
&\hspace{-2mm}=&\hspace{-2mm}z^2\bar{z}(e^{i\omega_0(-2\tau_1+\tau_2)}+2e^{-i\omega_0\tau_2})+\cdots.\nonumber\\
%*********************
u_t(-\tau_1)u_t^2(-\tau_2) &\hspace{-2mm}=&\hspace{-2mm} \big(w(z,\bar{z},\tau_1)+ze^{-i\omega_0\tau_1}+\bar{z}e^{i\omega_0\tau_1}\big)\nonumber\\
&\hspace{-2mm}&\hspace{-2mm}\times \big(w(z,\bar{z},\tau_2)+ze^{-i\omega_0\tau_2}+\bar{z}e^{i\omega_0\tau_2}\big)^2\nonumber\\
&\hspace{-2mm}=&\hspace{-2mm}z^2\bar{z}(e^{i\omega_0(-2\tau_2+\tau_1)}+2e^{-i\omega_0\tau_1})+\cdots.\nonumber
\end{eqnarray}
Using the above expressions, we can find the expression for other quadratic and cubic terms of $u_t$ by substituting appropriate values for $\tau_1$ and $\tau_2$. \\
Recall that
\begin{eqnarray}
   g(z,\bar{z}) &=& \bar{q}^*(0).\mathcal{F}_0(z,\bar{z})\nonumber\\
  g(z,\bar{z}) &=& g_{20}\frac{z^2}{2}+g_{11}z\bar{z}+g_{02}\frac{\bar{z}^2}{2}+g_{21}\frac{z^2\bar{z}}{2}+\cdots.\nonumber
\end{eqnarray}
Comparing the coefficients of $z^2,z\bar{z},\bar{z}^2$, and $z^2\bar{z}$, we get
\begin{eqnarray}
g_{20}=&&\hspace{-6mm}\bar{D}\kappa[2\xi_{xy}e^{-i\omega_0\tau_1}+2\xi_{xz}e^{-i\omega_0\tau_2}\nonumber\\
&&\hspace{-6mm}+\ 2\xi_{yy}e^{-2i\omega_0\tau_1}+2\xi_{yz}e^{-i\omega_0(\tau_1+\tau_2)}\nonumber\\
&&\hspace{-6mm}+\ 2\xi_{zz}e^{-2i\omega_0\tau_2}]\label{g20}\nonumber\\
 g_{11}=&&\hspace{-6mm}\bar{D}\kappa[\xi_{xy}(e^{-i\omega_0\tau_1}+e^{i\omega_0\tau_1})\nonumber\\
 &&\hspace{-6mm}+\ \xi_{xz}(e^{-i\omega_0\tau_2}+e^{i\omega_0\tau_2})+2\xi_{yy}\nonumber\\
 &&\hspace{-6mm}+\ \xi_{yz}(e^{-i\omega_0(\tau_1-\tau_2)}+e^{i\omega_0(\tau_1-\tau_2)})+ 2\xi_{zz}]\nonumber\\
 g_{02}=&&\hspace{-6mm}\bar{D}\kappa[2\xi_{xy}e^{i\omega_0\tau_1}+2\xi_{xz}e^{i\omega_0\tau_2}\nonumber\\
 &&\hspace{-6mm}+\ 2\xi_{yy}e^{2i\omega_0\tau_1}+2\xi_{yz}e^{i\omega_0(\tau_1+\tau_2)}\nonumber\\
 &&\hspace{-6mm}+\ 2\xi_{zz}e^{2i\omega_0\tau_2}]\nonumber.
 \end{eqnarray}
\begin{eqnarray}
g_{21}=&&\hspace{-6mm}\bar{D}\kappa[\xi_{xy}\big(2w_{11}(0)e^{-i\omega_0\tau_1}+ w_{20}(0)e^{i\omega_0\tau_1}\nonumber\\
%&&\hspace{-6mm}+\ \xi_{xy}\big(2w_{11}(0)e^{-i\omega_0\tau_1}+ w_{20}(0)e^{i\omega_0\tau_1}\nonumber\\
&&\hspace{-6mm}+2w_{11}(-\tau_1)+w_{20}(-\tau_1)\big)\nonumber\\
&&\hspace{-6mm}+\ \xi_{xz}\big(2w_{11}(0)e^{-i\omega_0\tau_2}+w_{20}(0)e^{i\omega_0\tau_2}\nonumber\\
&&\hspace{-6mm}+\ 2w_{11}(-\tau_2)+w_{20}(-\tau_2)\big)\nonumber\\
&&\hspace{-6mm}+\ \xi_{yy}\big(4w_{11}(-\tau_1)e^{-i\omega_0\tau_1}+2w_{20}(-\tau_1)e^{i\omega_0\tau_1}\big)\nonumber\\
&&\hspace{-6mm}+\ \xi_{yz}\big(2w_{11}(-\tau_1)e^{-i\omega_0\tau_2}+w_{20}(-\tau_1)e^{i\omega_0\tau_2}\nonumber\\
&&\hspace{-6mm}+\ 2w_{11}(-\tau_2)e^{-i\omega_0\tau_1}+w_{20}(-\tau_2)e^{i\omega_0\tau_1}\big)\nonumber\\
&&\hspace{-6mm}+\ \xi_{zz}\big(4w_{11}(-\tau_2)e^{-i\omega_0\tau_2}+ 2w_{20}(-\tau_2)e^{i\omega_0\tau_2}\big)\nonumber\\
&&\hspace{-6mm}+\ \xi_{xyy}(2e^{-2i\omega_0\tau_1}+4) \nonumber\\
&&\hspace{-6mm}+\ \xi_{xzz}(2e^{-2i\omega_0\tau_2}+4) \nonumber\\
&&\hspace{-6mm}+\ \xi_{yyz}(2e^{i\omega_0(-2\tau_1+\tau_2)}+4e^{-i\omega_0\tau_2})\nonumber\\ 
&&\hspace{-6mm}+\ \xi_{yzz}(2e^{i\omega_0(-2\tau_2+\tau_1)}+4e^{-i\omega_0\tau_1})\nonumber\\ 
&&\hspace{-6mm}+\ \xi_{xyz}(2e^{i\omega_0(\tau_1-\tau_2)}+2e^{-i\omega_0(\tau_1-\tau_2)}\nonumber\\ 
&&\hspace{-6mm}+\  2e^{-i\omega_0(\tau_1+\tau_2)})+6\xi_{yyy}e^{-i\omega_0\tau_1}\nonumber\\
&&\hspace{-6mm}+\ 6\xi_{zzz}e^{-i\omega_0\tau_2}].\label{g21}
\end{eqnarray}
% Substituting the values, we get\\
% \begin{eqnarray}
% g_{20}=&&\hspace{-6mm}-4j\bar{D}\kappa\xi_{xy}sin{(\omega_0\tau_1)}.\label{g20}\\
% g_{11}=&&\hspace{-6mm}0.\label{g11}\\
% g_{02}=&&\hspace{-6mm}4j\bar{D}\kappa\xi_{xy}sin{(\omega_0\tau_1)}..\label{g02}
% \end{eqnarray}
For $\theta \in [-\tau,0)$, we have
\begin{eqnarray*}
H(z,\bar{z},\theta) &&\hspace{-6mm}= -2Re\{\bar{q}^*(0).\mathcal{F}_0q(\theta)\}\\
&&\hspace{-6mm}=-g(z,\bar{z})q(\theta) -\bar{g}(z,\bar{z})\bar{q}(\theta)\\
&&\hspace{-6mm}=-\Big(g_{20}\frac{z^2}{2}+g_{11}z\bar{z} + g_{02}\frac{\bar{z}^2}{2}+\cdots\Big)q(\theta)\\
&&\hspace{-6mm} \ \ \ -\Big(\bar{g}_{20}\frac{\bar{z}^2}{2}+\bar{g}_{11}z\bar{z} + \bar{g}_{02}\frac{z^2}{2}+\cdots\Big)\bar{q}(\theta).
\end{eqnarray*}
Now using (\ref{H_equ}), we obtain
\begin{eqnarray*}
  H_{20}(\theta) &&\hspace{-6mm}= -g_{20}q(\theta)-\bar{g}_{20}\bar{q}\theta   \\
  H_{11}(\theta) &&\hspace{-6mm}= -g_{11}q(\theta)-\bar{g}_{11}\bar{q}\theta.
\end{eqnarray*}
From (\ref{eq:Atheta}), (\ref{eq1}) and (\ref{eq2}), we derive the following:
\begin{eqnarray*}
  w'_{20}(\theta) =&&\hspace{-6mm} 2i\omega_0w_{20}(\theta)+g_{20}q(\theta)+\bar{g}_{02}\bar{q}(\theta), \label{eq4}\\
  w'_{11}(\theta) =&&\hspace{-6mm} g_{11}q(\theta)+\bar{g}_{11}\bar{q}(\theta).\label{eq5}
\end{eqnarray*}
Solving the above differential equations yields
\begin{eqnarray}
w_{20}(\theta) &\hspace{-3mm}=&\hspace{-3mm} -\frac{g_{20}}{i\omega_0}q(0)e^{i\omega_0\theta} -\frac{\bar{g}_{02}}{3i\omega_0}\bar{q}(0)e^{-i\omega_0\theta}+Ee^{2i\omega_0\theta}\nonumber\\&&\label{w20}\\
w_{11}(\theta) &\hspace{-3mm}=&\hspace{-3mm} \frac{g_{11}}{i\omega_0}q(0)e^{i\omega_0\theta} -\frac{\bar{g}_{11}}{i\omega_0}\bar{q}(0)e^{-i\omega_0\theta}+F\label{w11}
\end{eqnarray}
for some $E$ and $F$.\\
For $\theta=0$, we get\\
%$H(z,\bar{z},0)= -2Re(\bar{q}^*.\mathcal{F}_0q(0))+\mathcal{F}_0,$
\begin{eqnarray}
  H(z,\bar{z},0)=&&\hspace{-6mm} -2Re(\bar{q}^*.\mathcal{F}_0q(0))+\mathcal{F}_0,\nonumber\\
  H_{20}(0) =&&\hspace{-6mm} -g_{20}q(0) -\bar{g}_{20}\bar{q}(0)\nonumber\\
  &&\hspace{-6mm}+\kappa\big[2\xi_{xy}e^{-i\omega_0\tau_1}+2\xi_{xz}e^{-i\omega_0\tau_2}\nonumber\\
  &&\hspace{-6mm}+\ 2\xi_{yy}e^{-2i\omega_0\tau_1}+2\xi_{yz}e^{-i\omega_0(\tau_1+\tau_2)}\nonumber\\
  &&\hspace{-6mm}+\ 2\xi_{zz}e^{-2i\omega_0\tau_2}\big]\label{H_20}\\
  H_{11}(0) =&&\hspace{-6mm} -g_{11}q(0) -\bar{g}_{11}\bar{q}(0)\nonumber \nonumber\\
  &&\hspace{-6mm}+\kappa\big[\xi_{xy}(e^{-i\omega_0\tau_1}+e^{i\omega_0\tau_1})\nonumber\\
  &&\hspace{-6mm}+\ \xi_{xz}(e^{-i\omega_0\tau_2}+e^{i\omega_0\tau_2}) + 2\xi_{yy}\nonumber\\&&\hspace{-6mm}+  2\xi_{yz}(e^{-i\omega_0(\tau_1-\tau_2)}+e^{i\omega_0(\tau_1-\tau_2)})\nonumber\\&&\hspace{-6mm}+\ 2\xi_{zz}\big].\label{H_11}
\end{eqnarray}
Using (\ref{eq:Atheta}), (\ref{eq1}) and (\ref{eq2}), we get
\begin{eqnarray}
  &&\kappa\xi_yw_{20}(-\tau_1)+\kappa\xi_zw_{20}(-\tau_2)-2i\omega_0w_{20}(0)\nonumber\\
  &&=g_{20}q(0)+\bar{g}_{02}\bar{q}(0)\nonumber\\
  &&\qquad-\kappa\big[2\xi_{xy}e^{-i\omega_0\tau_1}+2\xi_{xz}e^{-i\omega_0\tau_2}\nonumber\\
  &&\qquad+\ 2\xi_{yy}e^{-2i\omega_0\tau_1}+2\xi_{yz}e^{-i\omega_0(\tau_1+\tau_2)}\nonumber\\&&\qquad+\ 2\xi_{zz}e^{-i\omega_0\tau_2}\big]\label{eq8} \\
  &&\kappa\xi_yw_{11}(-\tau_1)+\kappa\xi_zw_{11}(-\tau_2)\nonumber\\
  &&\quad=  g_{11}q(0)+\bar{g}_{11}\bar{q}(0)-\kappa\big[2\xi_{xy}e^{-i\omega_0\tau_1}\nonumber\\
  &&\qquad+\ 2\xi_{xz}e^{-i\omega_0\tau_2}+2\xi_{yy}e^{-2i\omega_0\tau_1}\nonumber\\
  &&\qquad+\ 2\xi_{yz}e^{-i\omega_0(\tau_1+\tau_2)}+2\xi_{zz}e^{-i\omega_0\tau_2}\big]. \label{eq9}
\end{eqnarray}
% We have the solution for $w_{20}(\theta)$ and $w_{11}(\theta)$ from (\ref{w20}) and (\ref{w11}) respectively. 
% Hence, evaluate $w_{11}(0)$, $w_{20}(0)$, $w_{11}(-\tau_1)$, $w_{20}(-\tau_1)$,  $w_{11}(-\tau_2)$, $w_{20}(-\tau_2)$, substitute into (\ref{eq8}) and (\ref{eq9}) respectively, and calculate $E,F$ as
Evaluate $w_{11}(0)$, $w_{20}(0)$, $w_{11}(-\tau_1)$, $w_{20}(-\tau_1)$,  $w_{11}(-\tau_2)$ and $w_{20}(-\tau_2)$ using (\ref{w20}) and (\ref{w11}), and substituting in (\ref{eq8}) and (\ref{eq9}), we get $E$ and $F$ as\\
\begin{eqnarray*}
  && E = \frac{-g_{20}}{\bar{D}(\kappa\xi_ye^{-2i\omega_0\tau_1}+\kappa\xi_ze^{-2i\omega_0\tau_2}-2i\omega_0)}, \\
&& F = \frac{-g_{11}}{\bar{D}\kappa(\xi_y+\xi_z)}.
\end{eqnarray*}
% where
% \begin{eqnarray}
% % \nonumber to remove numbering (before each equation)
%   \Phi_1 =&&\hspace{-6mm}(-2i\omega_0)\left(\frac{g_{20}}{i\omega_0}+\frac{\bar{g}_{02}}{3i\omega_0}\right) +\ \kappa\xi_y\left(\frac{g_{20}}{i\omega_0}e^{-i\omega_0\tau_1}\right.\nonumber\\
%   &&\hspace{-6mm}\left.+\ \frac{\bar{g}_{02}}{3i\omega_0}e^{i\omega_0\tau_1}\right) +\kappa\xi_z\left(\frac{g_{20}}{i\omega_0}e^{-i\omega_0\tau_2}\right.\nonumber\\
%   &&\hspace{-6mm}\left.+\frac{\bar{g}_{02}}{3i\omega_0}e^{i\omega_0\tau_2}\right)-H_{20}(0)\nonumber\\
%   \Phi_2 =&& \hspace{-6mm}-\ \kappa\xi_y\left(\frac{g_{11}}{i\omega_0}e^{-i\omega_0\tau_1}\right.\nonumber\\
%   &&\hspace{-6mm}\left.-\ \frac{\bar{g}_{11}}{i\omega_0}e^{i\omega_0\tau_1}\right)- \kappa\xi_z\left(\frac{g_{11}}{i\omega_0}e^{-i\omega_0\tau_2}\right.\nonumber\\
%   &&\hspace{-6mm}\left.-\ \frac{\bar{g}_{11}}{i\omega_0}e^{i\omega_0\tau_2}\right)-H_{11}(0)\nonumber.
% \end{eqnarray}

Thus the stability analysis of the Hopf bifurcation can now be performed using \cite{hassard1981}.
The quantities required to study the nature of the Hopf bifurcation are as follows\\
\begin{eqnarray}
\mu_2 &&\hspace{-6mm}= \dfrac{-\operatorname{Re}[c_1(0)]}{\alpha'(0)},\quad \beta_2= 2\operatorname{Re}[c_1(0)],\nonumber
%\beta &&\hspace{-6mm}= \epsilon^2\beta_2+\mathcal{O}(\epsilon^4)\quad \beta_2 = 2\operatorname{Re}[c_1(0)]\quad \epsilon = \sqrt{\frac{\mu}{\mu_2}}\nonumber\\\label{49}
\end{eqnarray}
where
\begin{eqnarray*}
c_1(0) &&\hspace{-6mm}= \dfrac{i}{2\omega_0}\left(g_{20}g_{11}-2|g_{11}|^2-\dfrac{1}{3}|g_{02}|^2\right)+\dfrac{g_{21}}{2},\nonumber\label{c10}\\
\alpha'(0) &&\hspace{-6mm}=\mathbf{Re}\left(\dfrac{d\lambda}{d\kappa}\right)_{\kappa=\kappa_c}.
\end{eqnarray*}
% $\alpha'(0)$ is the real component of $(d\lambda/d\kappa)$ evaluated at $\kappa = \kappa_c$
The direction and stability of Hopf bifurcation is determined by the sign of $\mu_2$ and  $\beta_2$ respectively.
If $\mu_2 >0(\mu_2 <0)$ then the Hopf bifurcation is \emph{super-critical(sub-critical)}. Simillarly, the bifurcating solutions are \emph{asymptotically orbitally stable(unstable)} if $\beta_2<0(\beta_2>0)$.\\

\end{document}